\documentclass[11pt,a4paper]{article}

\pdfoutput=1
\usepackage[usenames,dvipsnames]{color}
\usepackage{jcappub_accepted}
\usepackage{multicol}
\usepackage{amssymb}
\usepackage{amsmath}
\usepackage{graphicx}
\usepackage{afterpage}

\newcommand{\params}{\psi}
\newcommand{\Eobsi}{N_i}
\newcommand{\phiobsi}{\phi'_i}
\newcommand{\Etruei}{E_i}
\newcommand{\phitruei}{\phi_{i}}

\newcommand{\Like}{\mathcal{L}}
\newcommand{\ntot}{{n_\mathrm{tot}}}

\newcommand{\diff}{\mathrm{d}}
\newcommand{\data}{D}
\newcommand{\iclike}{\textsf{DarkSUSY\ 5.0.6}}

\def\urltilda{\kern -.15em\lower .7ex\hbox{\~{}}\kern .04em}

\title{Use of event-level neutrino telescope data in global fits for theories of new physics}

\author[1,a]{P.~Scott,}
\author[2,a]{C.~Savage,}
\author[2]{J.~Edsj\"o}
\author{and\vspace{2.5mm}\\}
\author{\hspace{-3.7mm} The IceCube Collaboration:\\
R.~Abbasi$^{3}$,
Y.~Abdou$^{4}$,
M.~Ackermann$^{5}$,
J.~Adams$^{6}$,
J.~A.~Aguilar$^{7}$,
M.~Ahlers$^{3}$,
D.~Altmann$^{8}$
% ...
}
\affiliation[a]{Corresponding authors: P.~Scott, M.~Danninger, C.~Savage}
\affiliation[1]{Dept.~of Physics, McGill University, 3600 rue University, Montr\'eal, QC, H3A 2T8, Canada}
\affiliation[2]{Oskar Klein Centre for Cosmoparticle Physics and Dept.~of Physics, Stockholm University, SE-10691 Stockholm, Sweden}
\affiliation[3]{Dept.~of Physics and Wisconsin IceCube Particle Astrophysics Center, University of Wisconsin, Madison, WI 53706, USA}
\affiliation[4]{Dept.~of Physics and Astronomy, University of Gent, B-9000 Gent, Belgium}
\affiliation[5]{DESY, D-15735 Zeuthen, Germany}
\affiliation[6]{Dept.~of Physics and Astronomy, University of Canterbury, Private Bag 4800, Christchurch, New Zealand}
\affiliation[7]{D\'epartement de physique nucl\'eaire et corpusculaire, Universit\'e de Gen\`eve, CH-1211 Gen\`eve, Switzerland}
\affiliation[8]{Institut f\"ur Physik, Humboldt-Universit\"at zu Berlin, D-12489 Berlin, Germany}
% ...
\emailAdd{patscott@physics.mcgill.ca}
\emailAdd{danning@fysik.su.se}
\emailAdd{savage@fysik.su.se}

\abstract{
We present a fast likelihood method for including event-level neutrino telescope data in parameter explorations of theories for new physics, and announce its public release as part of \iclike.  Our construction includes both angular and spectral information about neutrino events, as well as their total number.  We also present a corresponding measure for simple model exclusion, which can be used for single models without reference to the rest of a parameter space.  We perform a number of supersymmetric parameter scans with IceCube data to illustrate the utility of the method: example global fits and a signal recovery in the constrained minimal supersymmetric standard model (CMSSM), and a model exclusion exercise in a 7-parameter phenomenological version of the MSSM.  The final IceCube detector configuration will probe almost the entire focus-point region of the CMSSM, as well as a number of MSSM-7 models that will not otherwise be accessible to e.g. direct detection.  Our method accurately recovers the mock signal, and provides tight constraints on model parameters and derived quantities.  We show that the inclusion of spectral information significantly improves the accuracy of the recovery, providing motivation for its use in future IceCube analyses.  
}

\keywords{dark matter theory, dark matter experiments, neutrino astronomy, cosmology of theories beyond the SM}

\begin{document} 

\maketitle

\section{Introduction}
\label{intro}

Despite ongoing efforts, we have yet to identify dark matter.  One of the most promising candidate classes is the weakly-interacting massive particle (WIMP) \cite{Bergstrom00,Bertone05,BertoneBook}, so named because WIMPs interact with standard model (SM) particles only via the weak nuclear force.  WIMPs are appealing because the expected cosmological density of a thermal relic with a weak-scale annihilation cross-section is the same order of magnitude as the observed value \cite{WMAP7}.

WIMPs are typically sought via three observational channels: direct WIMP-nucleon scattering (e.g.~\cite{CerdenoGreen10,Pato11,CoGeNTAnnMod11,CRESST11,XENON100,CDMSLowE12,COUPP12}), production at accelerators (e.g.~\cite{White07,Bai10,Goodman10,Fox12a,Fox12b,CMS12a,Frandsen12}) and indirect detection of SM products of WIMP self-annihilation (e.g.~\cite{Pamelapositron,CompositeAlt,LATDwarfComposite,MAGICSegue,VERITASSegue,IceCube09,IceCube09_KK}).  A promising version of indirect detection is to use neutrino telescopes such as IceCube \cite{IC40DM}, ANTARES \cite{Lim09} and SuperKamiokande \cite{SuperK11} to search for high-energy neutrinos produced by annihilation of WIMPs in the core of the Sun or Earth \cite{Press85,Krauss:1985ApJ,Freese86,Krauss86,Gaisser86}.  Due to their weak interactions with nuclei, such WIMPs would have scattered on solar nuclei and become gravitationally bound to the solar system, eventually returning to scatter repeatedly and settle (in the case of the Sun) to the solar core \cite{Gould87b} (see also \cite{Scott09} for a review of this process and impacts of WIMPs on stellar structure, and \cite{Gould91, Lundberg04, Peter09a, Peter09, Sivertsson12} for additional corrections due to the influence of planets).

WIMPs arise in many proposed extensions to the SM.  Standard neutrino telescope analyses \cite{IceCube09,IceCube09_KK,ICRC2011ic86,IC40DM,Lim09,SuperK11,Kumar12} take an effective view of WIMP interactions, placing limits on WIMP-nucleon scattering cross-sections as a function of the WIMP mass, by assuming a certain annihilation cross-section and final state.  These limits are difficult to translate into actual particle models, where the annihilation cross-section and branching fractions may take on a range of different values for any given WIMP mass and nuclear-scattering cross-section.  To properly interpret limits on neutrino fluxes in terms of the parameters (defined at e.g. the Lagrangian level) of a theory for new physics, it becomes necessary to compare the observed neutrino flux with the predicted neutrino signal for each individual point in the parameter space of the theory.  

Having translated the flux limits into direct constraints on the parameters of a theory, it is then also possible to compare and combine the sensitivities of multiple experiments, even if they probe entirely different sectors of the theory (e.g. neutrino and accelerator searches).  Within the realm of theories for new weak-scale physics, this `global fit' approach has so far been applied mostly to supersymmetry (SUSY) \cite{Baltz04,SFitter,Allanach06,Ruiz06,Trotta08,Fittino,Abdussalam09a, Abdussalam09b,Mastercode12}, in the form of the minimal supersymmetric standard model (MSSM).  Recent analyses have included new data from the Large Hadron Collider (LHC) \cite{MastercodeHiggs,SuperBayeSHiggs,Gfitter11}, direct \cite{BertoneLHCDD,Akrami11DD,MastercodeXENON100,SuperBayeSXENON100} and indirect detection \cite{Scott09c,Ripken11,Fittino12}.  Whilst great care needs to be taken over the details of the statistical methods employed \cite{Akrami09,SBSpike,SBcoverage,Akrami11coverage,Strege12}, these analyses have proven a clear success, pointing the way to a future of closer comparison between astronomical and terrestrial experiments.

To date global fits have not included neutrino telescope data.  The ability of future incarnations of IceCube to detect WIMP annihilation in the constrained MSSM (CMSSM) has been studied \cite{Trotta09,Nightmare,Roszkowski12}, based on the number of observed events only and simple estimates of the instrumental sensitivity.  Other authors have looked at predicted neutrino rates in 2D slices through MSSM parameter spaces (e.g. \cite{Feng01,Barger02,Baer04,Ellis09,Ellis11}), or from sets of models in non-statistical random scans of various incarnations of the MSSM (e.g. \cite{Bergstrom98b,Wikstrom09,Cotta12}).

Here we present a fast likelihood method for including full event-level data from neutrino telescopes in global fits and related analyses.  In particular, our formulation allows the directions and energy estimators associated with each event to be included in the final unbinned likelihood calculation, which can then be employed as a likelihood component in a global fit.  The inclusion of spectral information in neutrino searches for WIMPs has been of particular interest recently \cite{Rott11,Allahverdi12}.  As a by-product of our approach, we also present a rigorous but simple exclusion measure for individual models, based just on the observed number of events in a neutrino telescope.  We give full details of the input data required, and explicit examples using real data from the IceCube neutrino telescope.  

The IceCube data and simulations we employ are described in Section \ref{icecube}.  These data have been made publicly available on the IceCube webserver \cite{filesloc} and in \iclike.\footnote{\href{www.darksusy.org}{www.darksusy.org}}  The likelihood formalism, outlined in Section \ref{likelihood}, is implemented and also available in \iclike. Our example global fits and MSSM scans are detailed in Section \ref{examples}.  We summarise our method and examples in Section \ref{conclusions}.

\section{The IceCube Neutrino Telescope}
\label{icecube}

\subsection{Description}
\label{icdescription}

Completed on December 18 2010, the IceCube neutrino observatory~\cite{perfomancePaperIC} has $5160$ digital optical modules (DOMs) installed on $86$ strings between $1450$\,m and $2450$\,m below the surface in the glacial ice at the South Pole.  The detector consists of a hexagonal grid with horizontal spacing between strings of $125$\,m and a vertical spacing between DOMs of $17$\,m, leading to a total instrumented volume of $1$ km$^{3}$.  Eight of the 86 strings were deployed in a more densely instrumented core in the middle of the array, with average inter-string separation of 42\,m and vertical DOM separation of 7\,m.  Together with the 12 adjacent standard IceCube strings, this forms the DeepCore subarray.  DeepCore increases the sensitivity of IceCube at low energies, and substantially lowers the energy threshold.  IceCube records \v{C}erenkov light in the ice from relativistic charged particles created in neutrino interactions in or near the detector.  By recording the arrival times and locations of these photons with DOMs, the direction and energy of the muon, and identity of its parent neutrino, can be reconstructed.

\subsection{Data samples}
\label{icevent}

For this paper, we use the analysis details of a search for WIMP dark matter annihilation in the Sun with the IceCube $22$-string configuration~\cite{IceCube09,filesloc}.  This data set has $104.3$ days of live time, and was recorded between June 1 and September 23 $2007$. For the results in Sections~\ref{cmssmglobal} and~\ref{cmssmrecon} we use the event-level data at final analysis level and signal simulations from~\cite{IceCube09}. As described in Ref.~\cite{IceCube09}, final analysis level is reached through a series of increasingly stringent event selections that are applied to remove the cosmic-ray induced backgrounds. Events are selected which appear to originate from the direction of the Sun and which exhibit low energy signatures.

The results discussed in Section~\ref{mssm7} are based on a detailed study to determine the sensitivity of the $86$-string detector to signals originating from dark matter annihilation in the centre of the Sun~\cite{ICRC2011ic86}. This study was performed as a full analysis in all details and gives a realistic expectation of the capabilities of IceCube to observe dark matter-induced signals, given the state of data extraction, reconstruction, and signal discrimination techniques available at the time of the study.

\subsection{Signal and background simulation}
\label{icsim}

Previous work \cite{IceCube09,ICRC2011ic86} simulated solar WIMP signals using WimpSim~\cite{Blennow08,wimpsimweb}, which describes the annihilation of WIMPs inside the Sun. WimpSim simulates the production, interaction, oscillation and propagation of neutrinos from the core of the Sun to the detector, in a fully consistent three-flavour way. For the previous IceCube solar WIMP analyses that we consider (IceCube $22$-string data analysis \cite{IceCube09} and $86$-string sensitivity analysis \cite{ICRC2011ic86}), two annihilation channels were simulated: $\tilde{\chi}^{0}_{1}\tilde{\chi}^{0}_{1}\to W^{+}W^{-}$ (the `hard' channel) and $\tilde{\chi}^{0}_{1}\tilde{\chi}^{0}_{1}\to b\bar{b}$ (the `soft' channel).  These two channels were chosen in an attempt to approximately cover the range of possible SUSY models, by assuming 100\% branching into two channels with very different characteristics. For the $22$-string analysis, neutralino masses $m_{\chi^{0}_{1}}$ from $250$\,GeV to $5000$\,GeV were simulated. For the $86$-string analysis, this range was extended down to $m_{\chi^{0}_{1}} = 50$\,GeV.  A similar exercise was performed for Kaluza-Klein WIMPs in models of Universal Extra Dimensions \cite{IceCube09_KK}.

In contrast, the analysis method we describe here (Section~\ref{likelihood}) is developed to employ the specific annihilation spectrum appropriate for each model in a given theory of new physics.  We use various parameterisations of the MSSM as examples, but the analysis technique we present is applicable to any theory containing a WIMP.  To calculate neutrino annihilation spectra for individual SUSY models, we create yield tables from a series of WimpSim simulations, with neutralino masses from 3\,GeV to 10\,TeV and many different annihilation channels. For each SUSY model, we interpolate in these tables and add the contributions to the total neutrino spectrum from each annihilation channel, according to the partial annihilation cross section for the given channel. We include more complicated annihilation channels involving model-dependent decays, such as those of Higgs bosons, by summing up the contributions from their decay in flight. This procedure is included in DarkSUSY \cite{darksusy}, which we use to calculate the detailed neutrino spectra model by model.

Background contributions for this analysis are muon events from single and coincident air showers, as well as atmospheric neutrinos. No dedicated background simulations are needed for the work we present here, as we estimate the expected background from scrambling real data at the final analysis level (detailed within Section~\ref{icbackground}).

\subsection{Effective area calculation}
\label{iceffarea}
We derived detector responses for the 22-string and 86-string IceCube configurations mentioned in Section~\ref{icdescription} from the sets of signal simulations used in each analysis \cite{IceCube09,ICRC2011ic86}.  In each case, we calculated an effective area for detection of muon neutrinos by IceCube from the direction of the Sun as a function of neutrino energy.  These effective areas correspond to averages over the austral winter.  The 22-string area is available online \cite{filesloc}.  For convenience, both datafiles are also redistributed in \iclike.  The effective area for muon neutrinos and muon anti-neutrinos are given separately.  

Detector systematics determined from the analyses are energy-dependent; the data files therefore include systematic uncertainties on the detector response within each energy bin, to which we assign a $1\sigma$ confidence level. These uncertainties were determined within simulation studies, where identified sources of uncertainty, e.g. absolute DOM efficiency, photon propagation in ice or calibration constants, were individually varied within reasonable ranges of their original values.  Similarly, the uncertainties arising from limited simulation statistics are also given for each energy bin of the effective areas, at the $1\sigma$ confidence level.

\subsection{Angular response}
\label{icangular}

The point spread function (PSF) describes the uncertainty in the reconstructed arrival directions of incoming neutrinos.  The full PSF is a 2D joint probability distribution for the angular separation between the true arrival direction and the reconstructed one, in each of two linearly-independent directions on the sky.  For the small angles under consideration and suitably chosen axes, this is well approximated by a 2D Gaussian distribution of the form
\begin{equation}
\label{PSF2}
P(\theta_1,\theta_2) = \frac{1}{2\pi \sigma_1 \sigma_2} \, \exp\left[-\frac{\theta_1^2}{2\sigma_1^2}-\frac{\theta_2^2}{2\sigma_2^2}\right] \, ,
\end{equation}
where $\theta_1$ and $\theta_2$ are the angular separation along each axis.  Given that the Sun is a circularly symmetric source, we work exclusively with a reduction of the PSF into the 1-dimensional probability distribution for the absolute angular separation on the sky, $\phi\equiv\sqrt{\theta_1^2+\theta_2^2}$, removing the azimuthal dependence.  For an azimuthally-symmetric PSF, $\sigma = \sigma_1 = \sigma_2$ and the 1D PSF is 
\begin{equation}
\label{PSF}
P(\phi) = \frac{\phi}{\sigma^2} \, \exp\left[-\frac{\phi^2}{2\sigma^2}\right].
\end{equation}

Over a given energy range, it is possible to accurately construct the 1D PSF directly for neutrinos arriving from all directions, using IceCube data analysis signal simulations \cite{IceCube09,ICRC2011ic86}.  In that case, one need not assume that the 2D PSF is even Gaussian, let alone azimuthally symmetric.  However, the radial probability density function (PDF) of an azimuthally-symmetric 2D Gaussian is a reasonable approximation to the radial PDF of a moderately non-symmetric 2D Gaussian, and has the added advantage that it can be parameterised in terms of a single $\sigma$.  As the 2D PSF in IceCube is not strongly asymmetric, we therefore model the 1D PSF with Eq.~\ref{PSF}, but extract the parameter $\sigma$, which we refer to as the ``mean angular error'', directly from the 1D PSF constructed from IceCube signal simulations.  This can be done by integrating to the containment angle implied by e.g.\ the mean ($\langle\phi\rangle=\sqrt{\frac{\pi}{2}}\sigma$), second moment ($\langle\phi^2\rangle=2\sigma^2$), or median ($\sigma \sqrt{2\ln{2}}$; this is what we use here) of the $\phi$ distribution in the signal simulations.  As in previous work \cite{IceCube09,ICRC2011ic86}, we consider $\sigma$ over the same energy bins as used for calculating the effective area.

We associate angular uncertainties with real data events on an event-by-event basis, using the paraboloid method \cite{paraboloid}.  A paraboloid function is fitted to the muon track reconstruction likelihood function in the neighbourhood of the best fit.  The resulting confidence ellipse on the sky is represented by the axes $\sigma_{1}$ and $\sigma_{2}$, which correspond to the standard deviations of the likelihood function in each of two linearly-independent directions. The overall reconstructed likelihood track uncertainty, $\sigma_{\mathrm{para}}$ (the ``paraboloid sigma''), is calculated as the mean in quadrature of the uncertainties in the two axes, $\sigma_{\mathrm{para}}^2=(\sigma_{1}^{2} + \sigma_{2}^{2})/2$. Good track fits generally result in a paraboloid that is narrow along both axes, and therefore have small $\sigma_{\mathrm{para}}$ values.

The simulated distribution of reconstructed angles, used for determining the mean angular error, contains contributions from both the intrinsic angular deviation of neutrino-induced muons from the incoming neutrino direction, and the error in the reconstructed muon direction itself.  For actual data events, the former cannot be known except in an average sense, whereas the latter can be estimated for each event with the paraboloid sigma.  For the 22-string configuration, the reconstruction uncertainty dominates over the intrinsic angular deviation of neutrino-induced muons, so we are able to employ the paraboloid sigma $\sigma_{\mathrm{para}}$ as an accurate approximation to the total angular error in Eq.~\ref{PSF}, for individual events.  For future IceCube configurations or other experiments where the muon angular reconstruction is of higher precision than the intrinsic muon angular spread, an additional component must be added to $\sigma_{\mathrm{para}}$ for each event, in order to properly account for the difference between the incoming neutrino and neutrino-induced muon directions.

\begin{figure}[t]
\centering
\includegraphics[width=0.9\textwidth]{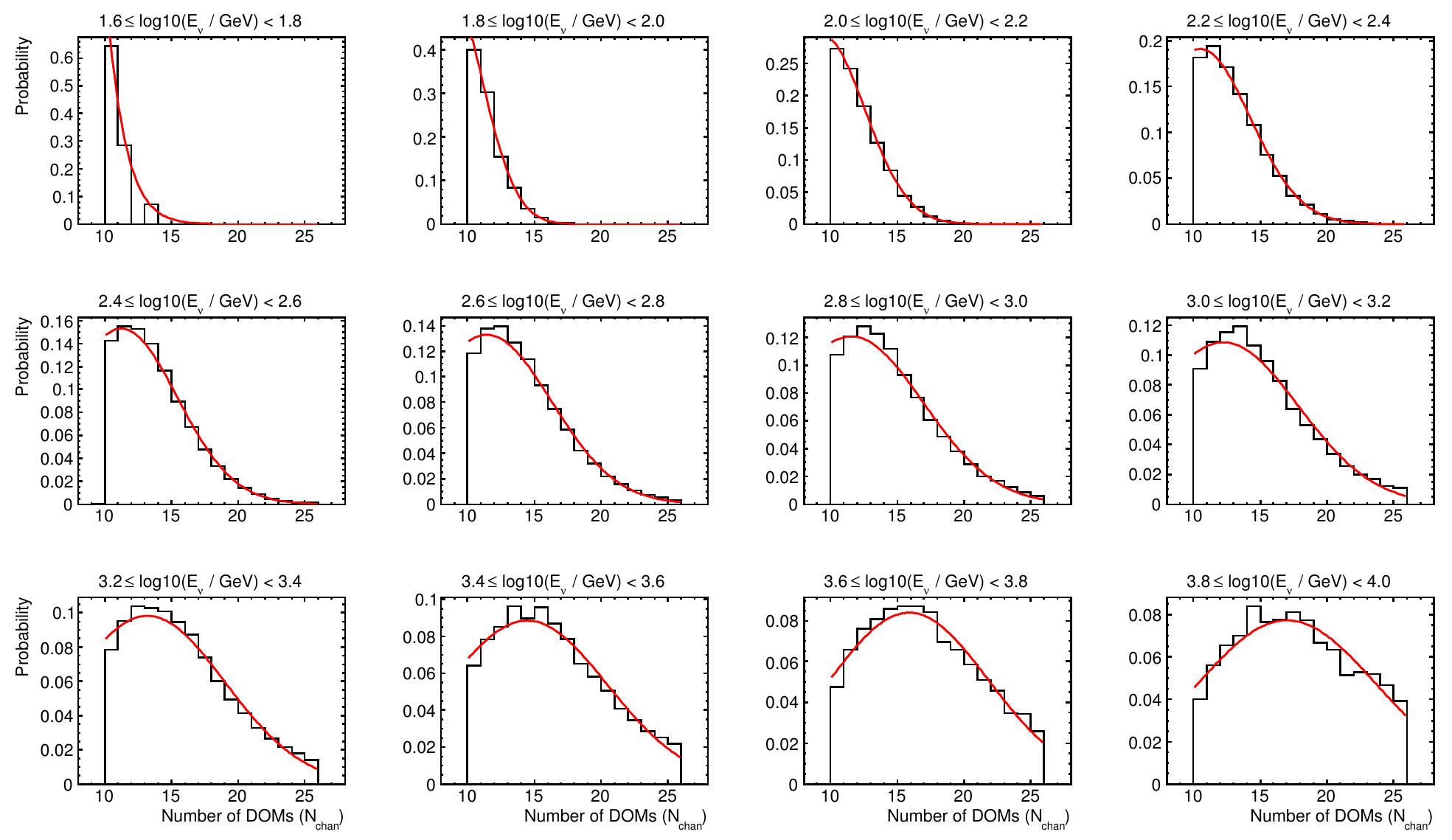}
\caption{Predicted probability distributions of $N_\mathrm{chan}$ derived from high-statistics $\nu$ simulations generated in the initial $22$-string analysis~\cite{IceCube09}.  Each distribution is constructed for neutrinos having energies in a specific logarithmic energy interval of width 0.2. The fitted Gaussian functions are to guide the eye only, and are not used in our calculations.}
\label{fig:nchan}
\end{figure}

\subsection{Energy estimator}
\label{icenergy}

In the broader context of IceCube data analysis, the study we present here deals with comparatively low-energy neutrinos. The corresponding muon events are associated with minimum-ionising tracks, where stochastic losses, which increase with muon energy, are not dominant. Energy estimators, as generally used within IceCube analyses targeting higher muon energies, exploit the energy-dependence of such losses in order to estimate the energy of the muon and the parent neutrino. The volume instrumented with the $22$-string configuration is too small to use the reconstructed track length of either fully or partially contained muon tracks to determine the energy. Instead, we use the number of lit DOMs $N_\mathrm{chan}$, a measure of the amount of recorded light per event, as a suitable energy estimator. We make no attempt to assign a specific energy and associated uncertainty to each event according to its $N_\mathrm{chan}$ value. Instead, we calculate the expected distribution of observed $N_\mathrm{chan}$ values for a series of intervals in neutrino energy, and use these together with the predicted energy spectrum of the signal for a given SUSY model, to calculate the predicted distribution of $N_\mathrm{chan}$.  Using an unbinned likelihood (Section~\ref{likelihood}), we then simply compare the observed value of $N_\mathrm{chan}$ for each event to this predicted distribution.  We derived the probability distributions per $\nu$-energy interval from high-statistics $\nu$ simulations generated in the initial $22$-string analysis \cite{IceCube09}.  These distributions are shown for the energy range relevant to our analysis in Fig.~\ref{fig:nchan}.  Note that we do not use the fitted Gaussian functions in Fig.~\ref{fig:nchan} for any purpose other than to guide the eye; we employ the actual distributions directly for our signal predictions and likelihood calculations.

%For future analyses of datasets taken with the $79$- or complete $86$-string array, we intend to use more accurate energy estimators, based on e.g. the reconstructed track length.

\subsection{Background estimation}
\label{icbackground}

We estimate the necessary background distributions from actual data.  For the likelihood functions in Section~\ref{likelihood}, we need two background distributions: the angular distribution of background events $\diff P_\mathrm{BG}(\phi')/\diff \phi'$ --- given as a function of $\phi'$, the angle between the reconstructed track direction and the Sun --- and the distribution of $N_\mathrm{chan}$ due to background events, $\diff P_\mathrm{BG}(N_\mathrm{chan})/\diff N_\mathrm{chan}$.

The angular distribution of the dominant cosmic-ray shower background contribution is expected to be azimuthally independent, depending only upon the angle from the horizon (zenith angle).  Thus, events incident in a horizontal band across the sky covering the same vertical range as the analysis (centred upon the solar zenith), but including all 360$^{\circ}$ around the horizon, are representative of the backgrounds in the small angular region around the Sun actually used in the analysis.  This allows for an estimate of the necessary background distributions driven purely by observation.\footnote{Technically, the 360$^{\circ}$ band includes the direction of the Sun, so the events in this band may include some signal as well as background.  However, the background overwhelmingly dominates any potential signal, so background estimates would be only negligibly affected.  If this was a concern, a section of the band including the Sun can be removed and only an e.g.\ 300$^{\circ}$ band used.}  The background is roughly constant per unit angular area in this band, with any mild zenith angle dependence softened by an averaging over the zenith angle of the Sun during the austral winter.  Hence $\frac{\diff P_{\mathrm{BG}}}{\diff\phi'} \sim \sin\phi'$, though the actual $\frac{\diff P_{\mathrm{BG}}}{\diff\phi'}$ we use is tabulated rather than fit to any functional form.

\subsection{Data format and availability}
\label{icdata}

The likelihood formalism described in the following Section, as implemented within \iclike, requires four simple text files containing IceCube data for any given analysis.  The first data file gives the binned $\nu$ and $\bar{\nu}$ effective areas as a function of energy, along with the associated $1\sigma$ statistical and systematic errors in each bin.  This file also provides the mean angular error. The second data file contains all events that a) passed the event selection corresponding to the effective areas, and b) have reconstructed directions that fall within some specified cut cone around the solar position.  The 22-string version of this file consists of 180 events within 10 degrees of the Sun.  This file includes the observed value of $N_\mathrm{chan}$ for each event, along with the reconstructed arrival angle relative to the Sun, and the corresponding paraboloid sigma.  This file also includes the total live (i.e.~exposure) time of observations towards the Sun. The third file gives the background distributions, with the angular distribution provided as a function of $\phi'$ and the spectral distribution as a function of $N_\mathrm{chan}$.  The final file contains the $N_\mathrm{chan}$ response histograms presented in Fig.~\ref{fig:nchan}.

We provide such files in \iclike\ for both the 22- and 86-string examples that we show in this paper.  The 22-string files (containing real data), are constructed from the original IceCube analysis \cite{IceCube09} and its associated data release \cite{filesloc}.  The 86-string files (containing simulated data), are constructed from the IceCube sensitivity analysis \cite{ICRC2011ic86} and the expected number of all-sky background events \cite{DanningerLicThesis}.\footnote{In both cases the effective areas have been updated slightly with respect to the published values, to properly account for neutrino-antineutrino asymmetries; these changes do not affect previous published IceCube limits.}  It is anticipated that 79-string data will later also be made available on the web and in DarkSUSY, in connection with an analysis of those data using the techniques presented in this paper.

\section{Likelihood functions}
\label{likelihood}

Using the data, responses and simulations described in Section~\ref{icecube}, our goal is to evaluate to what degree IceCube observations support or constrain a theory of new physics $\Psi$, given some particular values $\params$ of the theory's $m$ free parameters.  We refer to a specific new physics scenario $\Psi$ as a `theory' and a specific choice of its parameters $\params$ as a `model'.  We may be interested in
\begin{itemize}
\item[a)] the absolute goodness of fit of the model with parameter set $\params$, or
\item[b)] the goodness of fit of $\params$ relative to other points in the parameter space of the theory.
\end{itemize}
Case a) is model exclusion: we are interested in the maximal confidence level at which we can exclude the theory with specific parameter values $\params$ as a true hypothesis.  Case b) is most often parameter estimation: we are interested in determining how well $\params$ fits the data, compared to the best-fit point $\hat{\params}$ within the parameter space of the theory.  Assuming the overall theory to be correct, we use this information to determine a region of parameter space that includes the values of the parameters, with some specific level of confidence.

The absolute goodness of fit for a model is described by its $p$-value, where the confidence level with which the model can be excluded is $1-p$.  The $p$-value can be obtained from the likelihood function $\Like(\data|\params)$, where $\Like(\data|\params)\,\mathrm{d}\data$ describes the probability of observing data in the range $[\data, \data+\mathrm{d}\data]$ if $\Psi$ is true and the true values of its parameters are $\params$.  The relative goodness of fit can be obtained from the ratio 
\begin{equation}
\Lambda(\data|\params)\equiv\frac{\Like(\data|\params)}{\Like_\mathrm{max}(\data|\hat\params)}.
\end{equation}
Here $\Like_\mathrm{max}$ is the maximum likelihood, found at the best-fit point $\hat{\params}$.  Calculation of $p$ or $\Lambda$ requires knowledge of the distribution of $\Like(\data|\params)$ for repeated experiments (i.e.\ the distribution in data space). The simplest approach is to assume that $-2\ln\Lambda$ follows a $\chi^2$ distribution with $m$ degrees of freedom (as predicted in the limit of infinite data by Wilks' Theorem \cite{Wilks}), whilst the most rigorous approach is to explicitly construct the distribution by brute force simulation (the Feldman-Cousins approach \cite{FeldmanCousins}).

For both model exclusion and parameter estimation, we require an expression for $\Like(\data|\params)$. Here we develop a likelihood function for IceCube data, based on the total number of observed events and the individual properties of each event.  We go on to develop a $p$-value for model exclusion based on our likelihood construction.  Although we work exclusively at neutrino level in this paper (cf.\ Section~\ref{icevent}), the treatment in this Section should be equally valid at muon level.

\subsection{General unbinned likelihood}
Consider a set of $\ntot$ events observed in IceCube.  Denote the true energy of the $i$th event as $\Etruei$, and its arrival angle relative to the position of the Sun on the sky as $\phitruei$.  The $i$th event will be reconstructed with arrival angle $\phiobsi$, and will cause a number of DOMs $\Eobsi$ to fire.  The unbinned likelihood for these data is 
\begin{equation}
\label{unbinned_like}
\Like_\mathrm{unbin} \equiv \Like(\ntot|\theta_\mathrm{tot})\prod_{i=1}^{\ntot} \int_0^\pi\int_0^\infty Q(\Eobsi, \phiobsi | \Etruei, \phitruei)\frac{\diff P}{\diff \Etruei\,\diff \phitruei}(\Etruei,\phitruei,\params)\,\diff \Etruei\, \diff \phitruei.
\end{equation}
Here $Q(\Eobsi, \phiobsi | \Etruei, \phitruei)$ is the probability per unit angle and $N_\mathrm{chan}$ (i.e.\ probability density) for observing $\Eobsi$ and $\phiobsi$ for the $i$th event when the true values of the energy and arrival angle are $\Etruei$ and $\phitruei$.  The \textit{a priori} probability density for the $i$th event to arrive with energy $\Etruei$ from angle $\phitruei$ is given by $\frac{\diff P}{\diff \Etruei\,\diff \phitruei}(\Etruei,\phitruei,\params)$; this is a prediction of the model $\params$.  

\subsection{Number likelihood}
Neglecting systematic issues, which we address below, the prefactor $\Like(\ntot|\theta_\mathrm{tot})$ in Eq.~\ref{unbinned_like} is the standard Poissonian likelihood, equal to the probability of observing $\ntot$ events produced by a Poisson process with mean $\theta_\mathrm{tot}$
\begin{equation}
\label{standard}
\Like(\ntot|\theta_\mathrm{tot}) = \frac{\theta_\mathrm{tot}^\ntot e^{-\theta_\mathrm{tot}}}{\ntot!}.
\end{equation}
The expectation value for the number of events is $\theta_\mathrm{tot}\equiv\theta_\mathrm{tot}(\params)$, the total number of events predicted by model $\params$.  This is given by
\begin{equation}
\theta_\mathrm{tot}(\params) \equiv \theta_\mathrm{BG} + \theta_\mathrm{S}(\params),
\end{equation}
the sum of the predicted number of background events $\theta_\mathrm{BG}$ and signal events $\theta_\mathrm{S}(\params)$.  Whilst the predicted number of signal events depends on the model, the background does not.

Eq.~\ref{standard} accounts for statistical fluctuations in the number of observed events for a precisely-known mean $\theta_{\mathrm{tot}}$.  However, the predicted mean \textit{itself} may contain a systematic error, due to e.g.\ an error in the estimate of the effective area of the instrument.  Taking $\epsilon$ to be the (unknown) ratio of the true expected signal contribution $\theta_{\mathrm{S,true}}$ to the nominally-predicted signal contribution $\theta_{\mathrm{S}}$, we have $\theta_{\mathrm{S,true}} = \epsilon \theta_{\mathrm{S}}$.  This means that the relative (fractional) error on $\theta_{\mathrm{S}}$ is $\epsilon-1$.  To account for this potential systematic error in the number of predicted events, we marginalise over a probability distribution for the relative error $\epsilon-1$ in a semi-Bayesian manner \cite{Conrad03}.  Assuming a log-normal probability distribution for $\epsilon-1$, the likelihood becomes
\begin{equation}
\label{number}
\Like_\mathrm{num}(\ntot|\theta_\mathrm{tot}) = \frac{1}{\sqrt{2\pi}\sigma_\epsilon}\int_0^\infty \frac{(\theta_\mathrm{BG}+\epsilon\theta_\mathrm{S})^{\ntot}e^{-(\theta_\mathrm{BG}+\epsilon\theta_\mathrm{S})}}{\ntot!}\frac1\epsilon\exp\left[-\frac{1}{2}\left(\frac{\ln\epsilon}{\sigma_\epsilon}\right)^2\right]\mathrm{d}\epsilon \, .
\end{equation}
Here $\sigma_\epsilon$ is the fractional systematic error in the predicted number of events.  We take this term to be the sum in quadrature of the fractional uncertainty of the IceCube effective area, and a theoretical error $\tau$.  The fractional error in the effective area is itself the sum in quadrature of the fractional statistical and systematic errors determined in deriving the effective area (both have the character of a systematic for the purposes of the likelihood calculation, if not for the effective area calculation itself).  For $\tau$ we use a conservative error of 5\%, designed to account for neglected higher-order corrections and possible accumulated numerical round-off errors.

We apply the systematic uncertainty only to the \textit{signal} prediction in Eq.~\ref{number}, as neither an uncertainty in the signal prediction from theory, nor an error in the estimation of the effective area, would impact the number of predicted background events.  This is because $\theta_\mathrm{BG}$ is based on the total number of \textit{observed} events away from the Sun.  Eq.~\ref{number} gives the complete contribution of the number of events to the total likelihood; we hence refer to this as our adopted `number likelihood'.

We could have adopted a Gaussian probability distribution for $\epsilon-1$ in Eq.~\ref{number}, but this leads to a non-zero probability that $\epsilon=0$, which is unrealistic.  We have therefore adopted a log-normal distribution for $\epsilon$ in Eq.~\ref{number} and throughout this paper, although for small $\sigma_\epsilon$ the difference with the Gaussian case is minimal.  As the Gaussian calculation is significantly faster in some cases, we provide both Gaussian and log-normal routines in \iclike.

Eq.~\ref{number} assumes a single, constant systematic error on the effective area.  For real detectors, this error is typically highly energy-dependent.  In general there is no consistent way to allow the prior distribution for $\epsilon$ in Eq.~\ref{number} to vary with energy (or any other observable) in an unbinned likelihood calculation.  One option is to model the variation of the systematic error with energy using some parametric form, and then marginalise over the parameters; this however requires assumptions as to the functional form and the degree to which the size of the systematic is correlated at different energies.  Another solution is to bin events into broad `energy' ranges according to their observed $N_\mathrm{chan}$ values, using a Bayesian sorting technique that assumes a prior for the source spectral shape.  Each bin is then assigned a different systematic error for the effective area, based on the range of neutrino energies it covers, and an unbinned analysis is performed on the events \emph{within each bin}.  Ideally, one keeps the number of bins in such a setup to a minimum, to minimise prior-dependence in the final results.  A third, more conservative option, is simply to take $\sigma_\epsilon$ in Eq.~\ref{number} to be the largest percentage error seen on the effective area at any energy.  We follow the third strategy; we also attempted an analysis using the second, but found that the additional noise introduced by the need to sort events into the different bins grossly outweighed any advantage gleaned from having a more accurate effective area at higher neutrino energies.

\subsection{Spectral and angular likelihoods} 

In general $Q(\Eobsi, \phiobsi | \Etruei, \phitruei)$ in Eq.~\ref{unbinned_like} is a function of the spectral and angular resolution of the detector  
\begin{equation}
\label{sep_instrument}
Q(\Eobsi, \phiobsi | \Etruei, \phitruei,\params) = E_\mathrm{disp}(\Eobsi|\Etruei)PSF(\phiobsi|\phitruei), \\
\end{equation}
where $E_\mathrm{disp}$ is the distribution of $N_\mathrm{chan}$ for incoming neutrinos of energy $\Etruei$ (the energy dispersion of the instrument), and $PSF$ is the detector point-spread function.  Here we have assumed that $E_\mathrm{disp}$ has no angular dependence, and that the energy-dependence of the PSF can be neglected at this stage.  These assumptions are certainly not true in general, given the large width of $E_\mathrm{disp}$ and the strong energy-dependence of the PSF.  Here we apply a hard angular cut around the solar position, reducing the possible range of $\phitruei$, and therefore the range of variation of $E_\mathrm{disp}$ with angle.  Although this does not make $E_\mathrm{disp}$ entirely isotropic even within the analysis cone, our use of such a cone makes neglecting this angular variation reasonable.  The second approximation is well justified because our per-event analysis already takes the energy-dependence of the PSF into account implicitly, because we apply a unique PSF for each event using the paraboloid sigma (cf.~Section\ \ref{icangular}). 

Similarly, $\frac{\diff P}{\diff \Etruei\,\diff \phitruei}(\Etruei,\phitruei,\params)$ in Eq.~\ref{unbinned_like} depends on the true energy spectrum $\frac{\diff P}{\diff \Etruei}$ and spatial distribution $\frac{\diff P}{\diff \phitruei}$ of the signal and background
\begin{equation}
\label{sep_theory}
\frac{\diff P}{\diff \Etruei\,\diff \phitruei}(\Etruei,\phitruei) = \frac{\diff P}{\diff \Etruei}(\Etruei,\params) \frac{\diff P}{\diff \phitruei}(\phitruei,\params),
\end{equation}
with
\begin{align}
\label{dpde}
\frac{\diff P}{\diff \Etruei}(\Etruei,\params) &= f_\mathrm{BG}\frac{\diff P_\mathrm{BG}}{\diff \Etruei}(\Etruei) + f_\mathrm{S}\frac{\diff P_\mathrm{S}}{\diff \Etruei}(\Etruei,\params)\\
\label{dpdphi}
\frac{\diff P}{\diff \phitruei}(\phitruei,\params) &= f_\mathrm{BG}\frac{\diff P_\mathrm{BG}}{\diff \phitruei}(\phitruei)+f_\mathrm{S}\frac{\diff P_\mathrm{S}}{\diff \phitruei}(\phitruei,\params).
\end{align}
Here $\frac{\diff P_\mathrm{BG}}{\diff \Etruei}$, $\frac{\diff P_\mathrm{S}}{\diff \phitruei}$, etc refer to the signal and background contributions to the overall predicted spectral and angular distributions, and
\begin{equation}
f_\mathrm{S}=f_\mathrm{S}(\params)\equiv\frac{\theta_\mathrm{S}(\params)}{\theta_\mathrm{tot}(\params)}, \hspace{5mm} f_\mathrm{BG}=f_\mathrm{BG}(\params)\equiv\frac{\theta_\mathrm{BG}}{\theta_\mathrm{tot}(\params)}
\end{equation}
are the signal and background fractions, with $\theta_\mathrm{S}$ and $\theta_\mathrm{BG}$ the total predicted number of signal and background events within the analysis cone.  Like $\theta_\mathrm{tot}$ and $\theta_\mathrm{S}$, the fractions $f_\mathrm{S}$ and $f_\mathrm{BG}$ depend upon the model $\params$ by definition, but we typically do not write this dependence out explicitly.\footnote{Note that although not all the actual background is genuinely due to neutrinos (there is a large atmospheric muon component), we can simply interpret the parts of Eqs.~\ref{sep_theory}--\ref{dpdphi} due to non-neutrino events as effective neutrino angles and energies corresponding to those tracks.  As we only ever deal with the observed background distributions, not the predicted backgrounds (cf.~Eq.~\ref{anglike}), this is entirely valid.}

In Eqs.~\ref{sep_theory}--\ref{dpdphi} we have again dropped any explicit energy dependence in the angular part, or angular dependence in the spectral part.  This is perfectly well justified for the background component, as the arrival directions and $N_\mathrm{chan}$ values (and therefore by inference, energies) of background events are observed to be essentially uncorrelated.  For the signal component alone, the assumption that the energy spectrum does not vary with arrival angle is well justified because the IceCube PSF is far larger than the the angular extent of the region where high-energy neutrinos are produced in the Sun, regardless of the energies of the neutrinos.  That the angular extent of the signal does not depend on the energy of the neutrinos from the solar core, nor therefore the model $\params$, follows on exactly the same grounds.  (We can therefore model the angular distribution of the signal as simply a delta function at the solar position.)  

However, these assumptions do not rigorously hold for the \textit{combined} signal and background prediction, as the spatial and spectral characteristics of the signal and background differ; neutrinos coming from the Sun should presumably exhibit a more signal-like spectrum on average than those arriving from elsewhere on the sky.  We deal with this complication implicitly by way of our finite analysis cone.  We evaluate the signal spectral distribution in Eq.~\ref{dpde} at precisely the solar position, and then weight its contribution against that of the background by way of the total predicted signal and background fractions $f_\mathrm{S}$ and $f_\mathrm{BG}$.  These fractions are the integrated predictions \textit{over the full analysis cone}, meaning that in Eq.~\ref{dpde} we essentially derive a mean predicted energy spectrum over the full cone.  Similarly, we implicitly account for the energy-dependence of the angular distribution by weighting the signal and background angular distributions by the total signal and background fractions.

Collecting all the terms from Eq.~\ref{unbinned_like} where we have retained an explicit dependence on the energy of an incoming neutrino or the number of DOMs it triggers in the detector, we see that the contribution to the overall likelihood from the spectral properties (i.e.~$\Eobsi$) of event $i$ is   
\begin{align}
\label{energyintegral0}
\Like_{\mathrm{spec},i}(\Eobsi|\params)&\equiv\int_0^\infty E_\mathrm{disp}(\Eobsi|\Etruei) \frac{\diff P}{\diff \Etruei}(\Etruei,\params) \,\diff \Etruei \\
\label{energyintegral}
&= \int_0^\infty \left[f_\mathrm{BG}E_\mathrm{disp}(\Eobsi|\Etruei)\frac{\diff P_\mathrm{BG}}{\diff \Etruei}(\Etruei)+ f_\mathrm{S}E_\mathrm{disp}(\Eobsi|\Etruei)\frac{\diff P_\mathrm{S}}{\diff \Etruei}(\Etruei,\params) \right]\,\diff \Etruei.
\end{align}
The true energy spectrum of the atmospheric background $\frac{\diff P_\mathrm{BG}}{\diff \Etruei}$ would need to be estimated empirically using the observed background flux away from the Sun.  The distribution actually observed by IceCube (Section~\ref{icbackground}) is the true background distribution $\frac{\diff P_\mathrm{BG}}{\diff \Etruei}$ convolved with $E_\mathrm{disp}$; this is simply the first term in Eq.~\ref{energyintegral}, so
\begin{equation}
\label{speclike}
\Like_{\mathrm{spec},i}(\Eobsi|\params) = f_\mathrm{BG}\frac{\diff P_\mathrm{BG}}{\diff \Eobsi}(\Eobsi) +f_\mathrm{S}\int_0^\infty E_\mathrm{disp}(\Eobsi|\Etruei)\frac{\diff P_\mathrm{S}}{\diff \Etruei}(\Etruei,\params) \,\diff \Etruei.
\end{equation}

Collecting the remaining terms in Eq.~\ref{unbinned_like} with an explicit direction-dependence, the angular likelihood for event $i$ is
\begin{align}
\label{angularintegral0}
\Like_{\mathrm{ang},i}(\phiobsi|\params)&\equiv\int^\pi_0 PSF(\phiobsi|\phitruei)\frac{\diff P}{\diff \phitruei}(\phitruei,\params) \,\diff \phitruei \\
\label{angularintegral}
&=\int^\pi_0 f_\mathrm{BG}\frac{\diff P_\mathrm{BG}}{\diff \phitruei}(\phitruei)PSF(\phiobsi|\phitruei)+f_\mathrm{S}\frac{\diff P_\mathrm{S}}{\diff \phitruei}(\phitruei,\params)PSF(\phiobsi|\phitruei)\,\diff \phitruei.
\end{align}
As for the spectral distribution of the background, we estimate the angular dependence of the background empirically, based on observations away from the Sun (cf.\ Section~\ref{icbackground}).  This provides a direct estimate of the first term in Eq.~\ref{angularintegral}.

The predicted angular distribution of signal events $\frac{\diff P_\mathrm{S}}{\diff \phitruei}$ depends on the actual angular dependence of the signal.  Given that the Sun is a point source to IceCube, we can simply write $\diff P_\mathrm{S}/\diff \phitruei = \delta(\phitruei)$.\footnote{For an extended source, one also needs to take into account any anisotropy in the exposure time $t_\mathrm{exp}(\phi)$ of the detector, defined as the total time over which the detector is sensitive to neutrinos arriving from true angle $\phi$.  The effective exposure may vary across the sky due to the seasonal variation of the detector orientation, and choice of analysis cuts.  In this case, the predicted distribution would be
\begin{equation}
\label{signal_angular_dist}
\frac{\diff P_\mathrm{S}}{\diff \phitruei}(\phitruei) = \left.\frac{\diff P_\mathrm{S,i}}{\diff \phitruei}(\phitruei)t_\mathrm{exp}(\phitruei)\middle/\int_0^\pi\frac{\diff P_\mathrm{S,i}}{\diff \tilde\phitruei}(\tilde\phi) t_\mathrm{exp}(\tilde\phi)\,\diff\tilde\phi\right.,
\end{equation}
where $\diff P_\mathrm{S,i}/\diff \phitruei$ is the intrinsic source distribution, and the denominator simply ensures correct normalisation.}  
We can therefore easily solve the integral in the second term in Eq.~\ref{angularintegral}.  Combining this with the known angular distribution of background events, Eq.~\ref{angularintegral} becomes 
\begin{align}
\label{anglike}
\Like_{\mathrm{ang},i}(\phiobsi|\params) = f_\mathrm{S}PSF(\phiobsi|\phitruei=0) + f_\mathrm{BG}\frac{\diff P_\mathrm{BG}}{\diff \phiobsi}(\phiobsi).
\end{align}

$PSF(\phiobsi|\phitruei)$ in Eqs.~\ref{sep_instrument}, \ref{angularintegral0}, \ref{angularintegral} and \ref{anglike} refers to the unique 1D reduced PSF for each event $i$, as determined by its paraboloid sigma $\sigma_i$.  With a full PSF Gaussian in 2D (cf.\ Section~\ref{icangular}) and an analysis cone cut angle $\phi'_\mathrm{cut}$, the reduced PSF is given by
\begin{equation}
PSF(\phiobsi|\phitruei) = C(\phitruei,\sigma_i,\phi'_\mathrm{cut})\frac{|\phiobsi-\phitruei|}{\sigma_i^{\phantom{i}2}}\exp\left[-\frac12\left(\frac{\phiobsi-\phitruei}{\sigma_i}\right)^2\right],
\end{equation}
where
\begin{equation}
C(\phitruei,\sigma_i,\phi'_\mathrm{cut})^{-1} = 1 - \mathrm{exp}\left[-\frac12\left(\frac{\phi'_\mathrm{cut}-\phitruei}{\sigma_i}\right)^2\right]
\end{equation}
is an angular correction factor that ensures the PSF is unitary within the analysis cone (i.e. the integrated probability for any given event to have come from anywhere at all is 1).

\subsection{Total likelihood function}

To arrive at the final composite unbinned likelihood, we simply substitute Eqs.~\ref{number}, \ref{speclike} and \ref{anglike} into Eq.~\ref{unbinned_like}, giving  
\begin{equation}
\label{full_final_unbinned_like}
\Like_\mathrm{total}(\ntot,\Xi|\params) = \Like_\mathrm{num}(\ntot|\params) \prod_{i=1}^{\ntot}\Like_{\mathrm{ang},i}(\phiobsi|\params)\,\Like_{\mathrm{spec},i}(\Eobsi|\params),
\end{equation}
where $\Xi\equiv\{\Eobsi,\phiobsi\}_{i=1..\ntot}$ is the set of all reconstructed event arrival directions and $N_\mathrm{chan}$ values.  This is the likelihood function we employ for all the examples of parameter estimation in this paper.

\subsection{Predicted distributions and event counts}
\label{predicted}

So far we have focused on the construction of the likelihood function, assuming that one knows how to calculate the number of events and their distribution in energy predicted by a model $\params$.  The signal contains contributions from both neutrinos and anti-neutrinos,
\begin{equation}
\theta_\mathrm{S}(\params) \equiv \theta_{\mathrm{S},\nu}(\params) + \theta_{\mathrm{S},{\bar{\nu}}}(\params).
\end{equation}
The energy distribution of the predicted signal is the normalised sum of the neutrino and anti-neutrino flux spectra indicated by model $\params$, each weighted by the respective effective area of the detector ($A_\nu$ or $A_{\bar{\nu}}$).  The sum is further weighted by an overall angular loss factor $L$, giving
\begin{equation}
\label{spectral_shape}
\frac{\diff P_{\mathrm{S}}}{\diff E}(E,\params) = \frac{t_\mathrm{exp}(\phi=0)}{\theta_{\mathrm{S}}}L(E,\phi'_\mathrm{cut})\left[A_\nu(E)\frac{\diff \Phi_{\mathrm{S},\nu}}{\diff E}(E,\params)+A_{\bar{\nu}}(E)\frac{\diff \Phi_{\mathrm{S},\bar{\nu}}}{\diff E}(E,\params)\right].
\end{equation}
Here $\frac{\diff \Phi_{\mathrm{S},\nu}}{\diff E}(E,\params)$ and $\frac{\diff \Phi_{\mathrm{S},\bar{\nu}}}{\diff E}(E,\params)$ are the respective differential neutrino and anti-neutrino fluxes incident on the detector (due to neutrino production in the Sun), with units of events per unit area, energy and time.  The prefactor $t_\mathrm{exp}(\phi=0)/\theta_{\mathrm{S},\nu}$ ensures that Eq.~\ref{spectral_shape} is indeed normalised.  Both $A$ and $L$ are explicitly energy-dependent; $A$ has units of area and $L$ is dimensionless.

The angular loss factor $L(E,\phi'_\mathrm{cut})$ accounts for events that originate from the Sun, but are so badly reconstructed that they fall outside the analysis cut cone ($\phiobsi>\phi'_\mathrm{cut}$).  We use the (1D, reduced) mean PSF of the instrument to determine $L$, as defined by the energy-dependent mean angular error $\sigma(E)$ (cf.\ Section~\ref{icangular}).  This gives
\begin{equation}
\label{anglossfactor}
L(E,\phi'_\mathrm{cut}) = 1 - \mathrm{exp}\left[-\frac12\left(\frac{{\phi'_\mathrm{cut}}}{\sigma(E)}\right)^2\right].
\end{equation}

We can simply read off the definition of $\theta_{\mathrm{S},\nu}$ from Eq.~\ref{spectral_shape} by virtue of it being a normalised distribution
\begin{equation}
\label{thetaS}
\theta_{\mathrm{S},\nu}\equiv t_\mathrm{exp}(\phi=0)\int_0^\infty L(E)A_\nu(E)\frac{\diff \Phi_{\mathrm{S},\nu}}{\diff E}(E,\params) \diff E.
\end{equation}
The corresponding expression for $\theta_{\mathrm{S},{\bar{\nu}}}$ is Eq.~\ref{thetaS} with $\nu\rightarrow\bar{\nu}$.  This integral has no cutoff, as $\lim_{E\to0}A(E)=\lim_{E\to\infty}A(E)=0$.

The total number of predicted background events in the analysis cone ($\theta_\mathrm{BG}$) can be determined by simply rescaling the background rate observed away from the Sun, to the desired exposure time and analysis cone sky fraction.

\subsection{$p$-value for model exclusion}

Unfortunately, there is no established method for determining goodness-of-fit using unbinned maximum likelihood estimators, short of a full Neyman construction using Monte Carlo simulations based on Eq.~\ref{full_final_unbinned_like}.  Whilst this is possible for simple toy models, it is not a computationally feasible option for global fits in virtually any UV-complete theory; lengthy renormalisation-group and relic density calculations make it unrealistic for analyses of supersymmetry, for example.

A more realistic option is to do model exclusion based entirely on the number of observed events, using Eq.~\ref{number}.  This has the obvious disadvantage of discarding the spectral and angular information that go into Eq.~\ref{full_final_unbinned_like}, reducing our overall ability to exclude models compared to a full Neyman construction.  However, this strategy has the distinct advantage that we know exactly how Eq.~\ref{number} is distributed in data space, as it simply describes a Poisson process.  This knowledge is what allows us to calculate an absolute $p$-value for any model, completely independent of the value of the likelihood function elsewhere in a theory's parameter space, and exclude that model with confidence $1-p$.

We begin by identifying each $\psi$ as a specific signal hypothesis, and identifying $\psi$ plus the background distribution inferred from observations away from the Sun as a specific signal+background hypothesis $s+b$.  The $p$-value for a hypothesis is defined \cite{Junk,James} as the probability, in a repeat of the chosen experiment, of obtaining a test statistic at least as extreme as the one actually observed, if the hypothesis is true.  The test statistic with which we are probing the $s+b$ hypothesis is the likelihood ratio, 
\begin{equation}
\label{likelihoodratio}
X \equiv \frac{\Like_\mathrm{num}(\ntot|\theta_\mathrm{S}+\theta_\mathrm{BG})}{\Like_\mathrm{num}(\ntot|\theta_\mathrm{BG})}.
\end{equation}
$X$ is the ratio of the number likelihood (Eq.~\ref{number}) for the signal+background prediction to the number likelihood for a background-only prediction.  A more extreme result is one that gives a lower likelihood ratio (i.e. is less probable if $s+b$ is true).  We do not probe the hypothesis $\psi$ directly with Eqs.~\ref{number} and \ref{likelihoodratio}, as $\psi$ alone does not give a specific prediction for the total number of observed events.

Referring to Eqs.~\ref{standard}, \ref{number} and \ref{likelihoodratio} we see that $X$ is in fact a monotonically increasing function of the number of observed events $n_\mathrm{tot}$.  This means that the $p$-value for a hypothesis, as tested using the number likelihood ratio $X$, is the total probability of observing $n_\mathrm{tot}$ or less events if the hypothesis is true.  This is precisely the sum of the likelihood over all possible numbers of observed events less than or equal to $n_\mathrm{tot}$.  For the signal plus background hypothesis, this is
\begin{equation}
\label{sbpval}
p_{s+b} = \sum_{n=0}^{n_\mathrm{tot}}\Like_\mathrm{num}\left[n|\theta_\mathrm{tot}\left(\psi\right)\right] \,,
\end{equation}
and for the background-only hypothesis
\begin{equation}
\label{bgpval}
p_{b} = \sum_{n=0}^{n_\mathrm{tot}}\Like_\mathrm{num}\left(n|\theta_\mathrm{BG}\right) \,.
\end{equation}
Note that there is a small conceptual complication in the case of the background-only hypothesis: direct substitution of $\theta_\mathrm{S}=0$ into Eq.~\ref{likelihoodratio} causes the distribution of the test statistic to collapse to a delta function at $X=1$.  Eq.~\ref{bgpval} is therefore strictly valid only when Eq.~\ref{likelihoodratio} is taken in the limit $\theta_\mathrm{S}\rightarrow 0$, rather than evaluated at exactly $\theta_\mathrm{S}=0$.  For further explanation, we refer the interested reader to the original description \cite{CLs1,CLs2} of the $CL_s$ technique (of which our $p$-value is essentially an example), and the conditioned confidence interval technique from which the $CL_s$ method was generalised \cite{Zech}.

For a true Poisson likelihood given by Eq.~\ref{standard}, the sums in Eqs.~\ref{sbpval} and \ref{bgpval} can be performed relatively painlessly, but for the ``smeared'' Poisson likelihood given by Eq.~\ref{number} (the form we actually use\footnote{Strictly speaking, it is Eq.~\ref{standard} that we know the distribution of exactly, not Eq.~\ref{number}.  Given that Eq.~\ref{number} is simply a smeared form of Eq.~\ref{standard} however, the two distributions should be very similar.}), the sum is over terms each requiring a numerical integration, a potentially slow process.  We provide optimised routines for handling the numerical integration and summing that ensure calculations of the $p$-values are relatively quick.

In a classical sense $p_{s+b}$ can be regarded as the $p$-value for testing the hypothesis $\psi$ when the background hypothesis is known to be correct.  However, it is sensitive to statistical fluctuations in the actual observed background; a downward fluctuation in the background can cause $\psi$ to be excluded with an unreasonably high confidence level, for example.  To correct for such problems, we use the modified $p$-value as our estimate for the hypothesis $\psi$ alone, defined as \cite{Junk}
\begin{equation}
\label{finalpval}
p_{\psi} = \frac{p_{s+b}}{p_{b}} \, .
\end{equation}

\section{Example SUSY scans and global fits}
\label{examples}

As concrete examples of our likelihood formalism in action, we perform three different analyses of the impact of IceCube data on theories of new physics.  We examine two variants of the minimal supersymmetric standard model (MSSM): the constrained MSSM (CMSSM), and a 7-parameter, low-energy phenomenological parameterisation (the MSSM-7).  In the first analysis, we consider the impact of including real IceCube 22-string data on a CMSSM global fit, using the full likelihood function (Eq.~\ref{full_final_unbinned_like}).  In this first analysis, as a rough proxy for the full detector we also rescale the effective area of the 22-string analysis, and consider the resulting impact on the CMSSM.  In the second analysis, we assess the ability of our method to recover a hypothetical WIMP annihilation signal from the Sun in terms of the CMSSM, using the same rescaled 22-string detector.  In the third analysis, we give an example of a simple model exclusion exercise in the MSSM-7, using Eq.~\ref{finalpval} with the 86-string detector simulation.

\subsection{CMSSM global fit with 22-string data}
\label{cmssmglobal}

The CMSSM is a high-energy parameterisation of the MSSM.  It is defined in terms of the sign of the Higgs mixing parameter $\mu$, the ratio of up-type to down-type Higgs VEVs $\tan\beta$, the universal trilinear coupling $A_0$ and the universal mass parameters $m_0$ and $m_{1/2}$.  $\tan\beta$ is defined at the electroweak scale, whereas $A_0$, $m_0$ and $m_{1/2}$ are defined at the scale of grand unification.  The WIMP mass, given by the mass of the lightest neutralino $m_{\chi_1^0}$, is a derived quantity.  We refer the reader to Ref.~\cite{BaerTata} for full details.

In order to include IceCube 22-string data in a CMSSM global fit, we coupled \iclike\ to \textsf{SuperBayeS 1.5.1} \cite{Ruiz06,Trotta08,Trotta09} and included Eq.~\ref{full_final_unbinned_like} as an additional likelihood component in the global fit.  The total likelihood function we employ for scans in this section includes all components available in the public release of \textsf{SuperBayeS 1.5.1}: LEP limits on sparticle and Higgs masses, the anomalous magnetic moment of the muon, limits on rare processes like $b\to s\gamma$, additional $B$-physics observables, the dark matter relic density, and measurements of SM nuisance parameters (refer to \cite{Trotta08,Trotta09} for details).  We do not include more recent constraints from the LHC \cite{Allanach11a,Allanach11b,Mastercode12,MastercodeHiggs,SuperBayeSHiggs,Gfitter11}, XENON-100 \cite{MastercodeXENON100,SuperBayeSXENON100,Fittino12} or indirect detection \cite{Scott09c,Ripken11,Fittino12}, as they are not necessary for our simple illustrative purposes here.  Future analyses using the likelihood formalism we describe, where the focus is on physical interpretation rather than methods, should include such data.

We performed a scan of the CMSSM parameter space using MultiNest \cite{MultiNest1,MultiNest} with 4000 live points and a tolerance of 0.5.  These are settings appropriate for mapping the posterior PDF in a CMSSM scan, but not the profile likelihood \cite{Akrami09,SBSpike}.  The likelihood formalism we present here is of course equally well adapted for use in frequentist as Bayesian analyses, but obtaining correctly converged profile likelihoods takes roughly an order of magnitude more computing power than posterior PDFs.  We thus rely almost exclusively on posterior maps in this paper, as we are essentially only interested in the results of the scans for the sake of illustration, so the choice of posterior PDF or profile likelihood is fairly arbitrary.  One must also be concerned with correct coverage in presenting profile likelihood results \cite{Akrami11coverage,SBcoverage,Strege12}, an additional unnecessary complication for our current purposes. 

An important point however is that the full IceCube likelihood calculation takes a fraction of a second in nearly all cases, so it is not a significant bottleneck in most scans when compared to e.g. a relic density calculation.

We parsed and plotted the results of our \textsf{SuperBayeS} runs using \textsf{pippi} \cite{pippi}.  The parsing step requires binning samples and marginalising posterior PDFs or profiling the likelihood function.  In most cases, we sorted samples into 100 bins in each parameter/observable, and interpolated between those bins for display, at a resolution of 500 points per parameter/observable.  Plots of the predicted number of signal events in IceCube are the exception: in this case we used 130 bins, and interpolated to a resolution of 3000 points per observable direction, to better resolve the region of interest. 

\begin{figure}[t]
\centering
\hspace{-0.15\textwidth}
\begin{minipage}[t]{0.31\textwidth}
\centering
\includegraphics[height=1.05\linewidth]{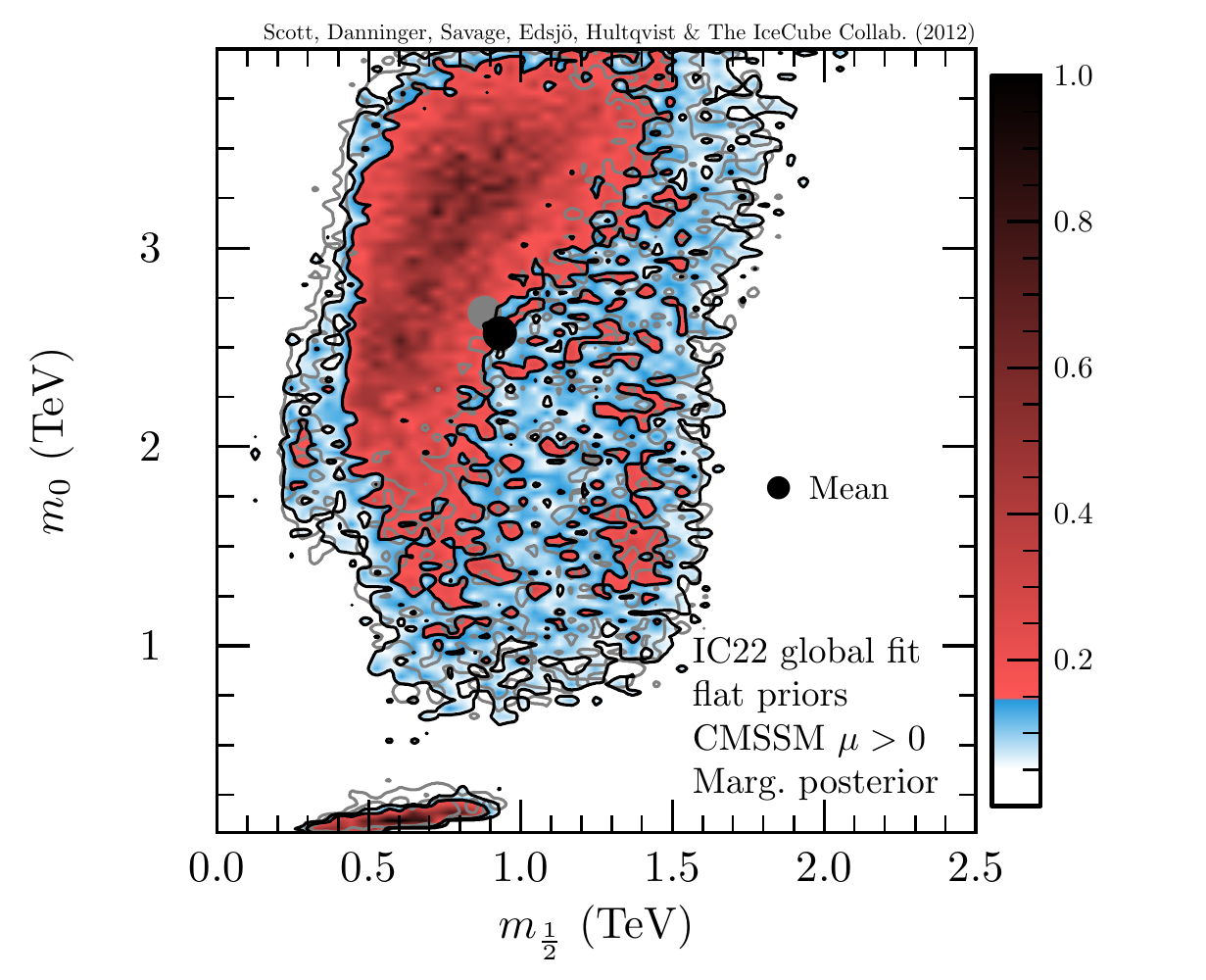}
\end{minipage}
\hspace{0.045\textwidth}
\begin{minipage}[t]{0.31\textwidth}
\centering
\includegraphics[height=1.05\linewidth]{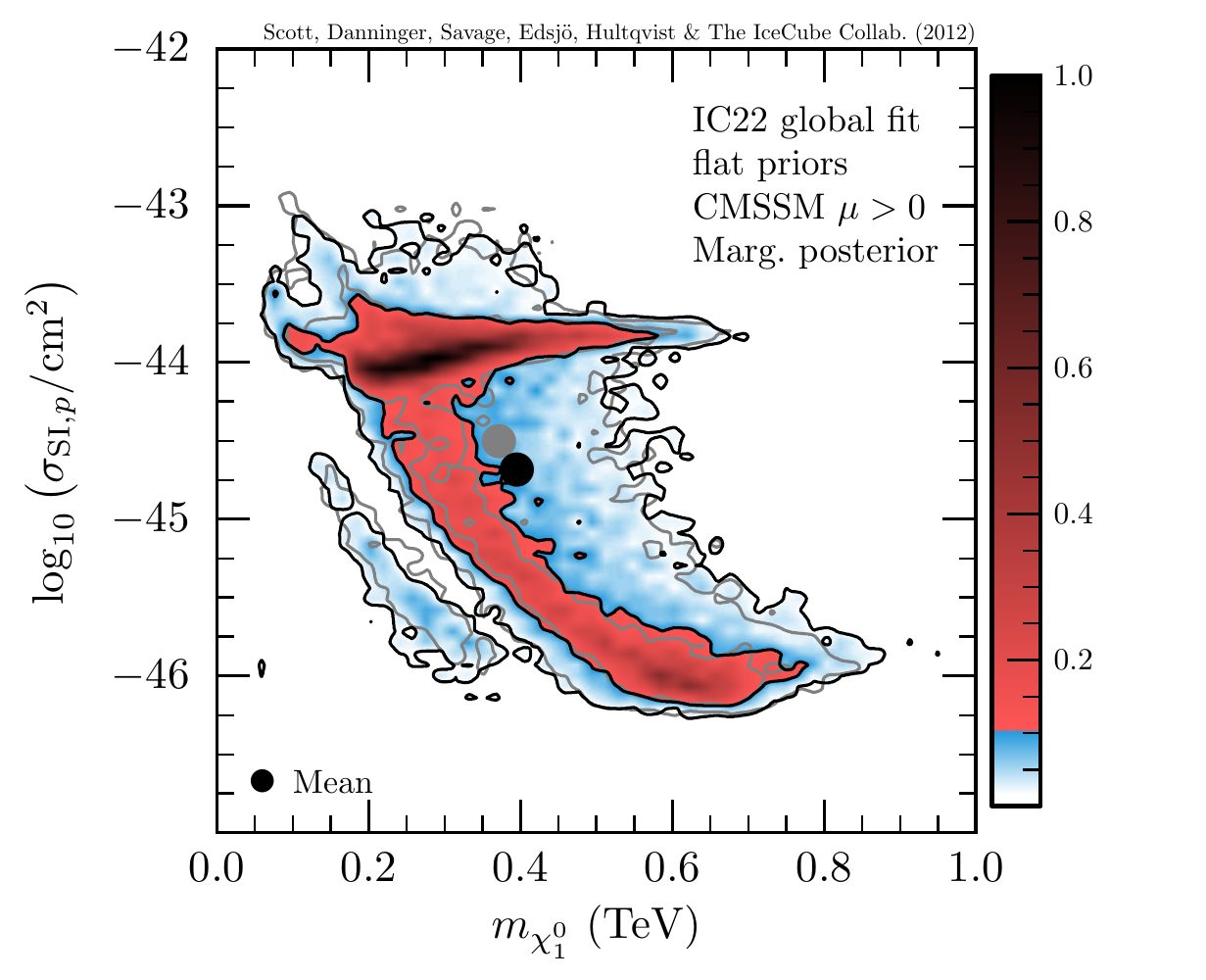}
\end{minipage}
\hspace{0.045\textwidth}
\begin{minipage}[t]{0.31\textwidth}
\centering
\includegraphics[height=1.05\linewidth]{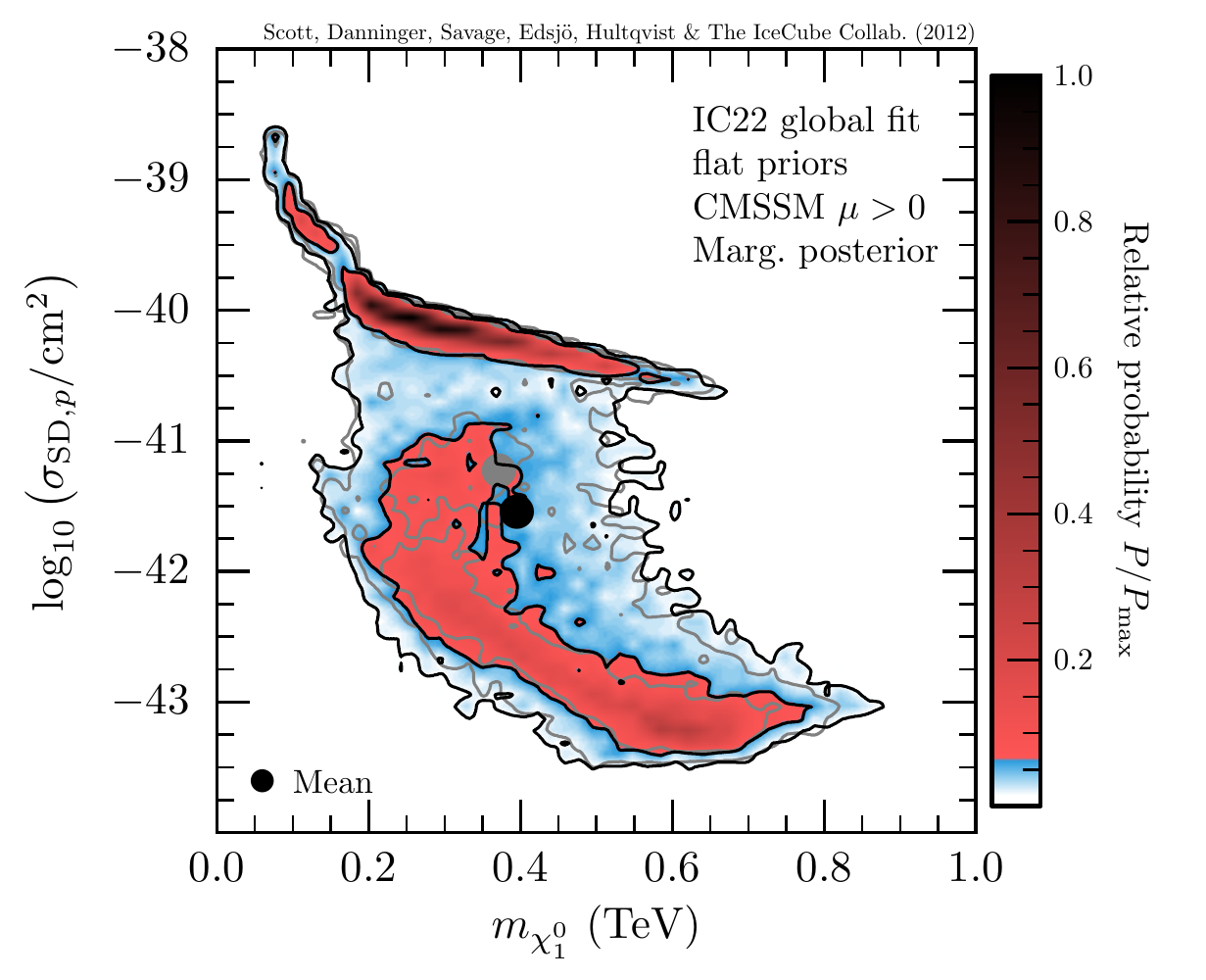}
\end{minipage}
\caption{Joint 2D posterior probability distributions from a CMSSM global fit including IceCube 22-string event data.  Contours indicate 68.3\% ($1\sigma$) and 95.4\% ($2\sigma$) credible regions.  Shading and black contours give the result for the fit with IceCube data, whilst grey contours correspond to an identical global fit, but with no IceCube data included.  Existing 22-string data has a minimal impact on the parameter space of the CMSSM.}
\label{fig:cmssmIC22}
\end{figure}

Our first global fit employed the observed events, detector response and effective area of the published 22-string (IC22) analysis \cite{IceCube09}.  For this analysis we used an angular cut of $\phi_\mathrm{cut}=10^\circ$ around the solar position.  The resulting marginalised posterior PDFs are are shown in colour in Fig.~\ref{fig:cmssmIC22}, along with contours indicating 68.3\% ($1\sigma$) and 95.4\% ($2\sigma$) credible intervals.  Here we show results in the standard $m_0$, $m_{1/2}$ parameter plane, as well as in terms of the neutralino mass -- nuclear scattering cross-section planes, for both spin-dependent and spin-independent interactions.  We also plot corresponding credible contours in grey for an identical scan run \textit{without} the inclusion of any IceCube data.  

\begin{figure}[t]
\centering
\hspace{-0.15\textwidth}
\begin{minipage}[t]{0.31\textwidth}
\centering
\includegraphics[height=1.05\linewidth]{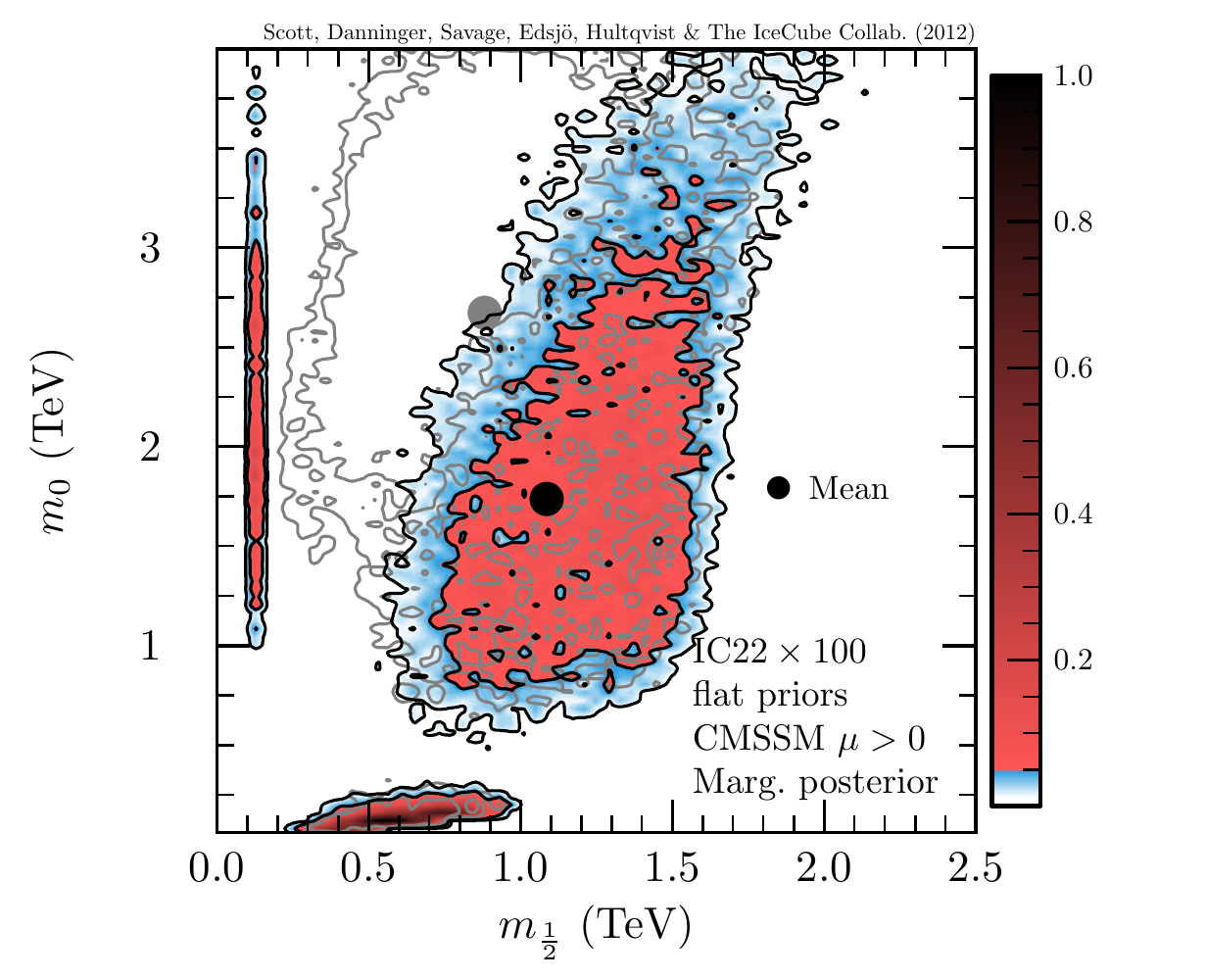}
\end{minipage}
\hspace{0.045\textwidth}
\begin{minipage}[t]{0.31\textwidth}
\centering
\includegraphics[height=1.05\linewidth]{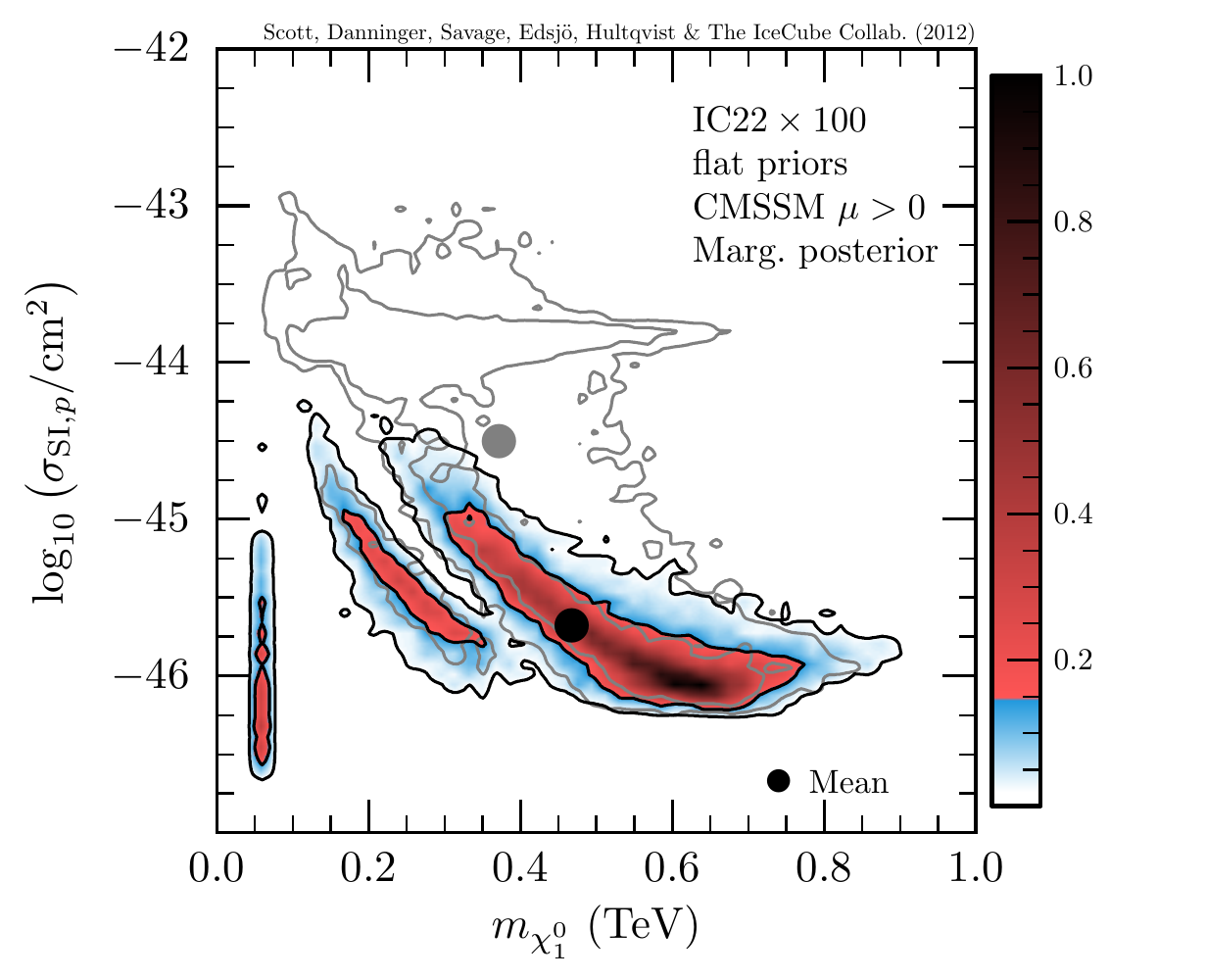}
\end{minipage}
\hspace{0.045\textwidth}
\begin{minipage}[t]{0.31\textwidth}
\centering
\includegraphics[height=1.05\linewidth]{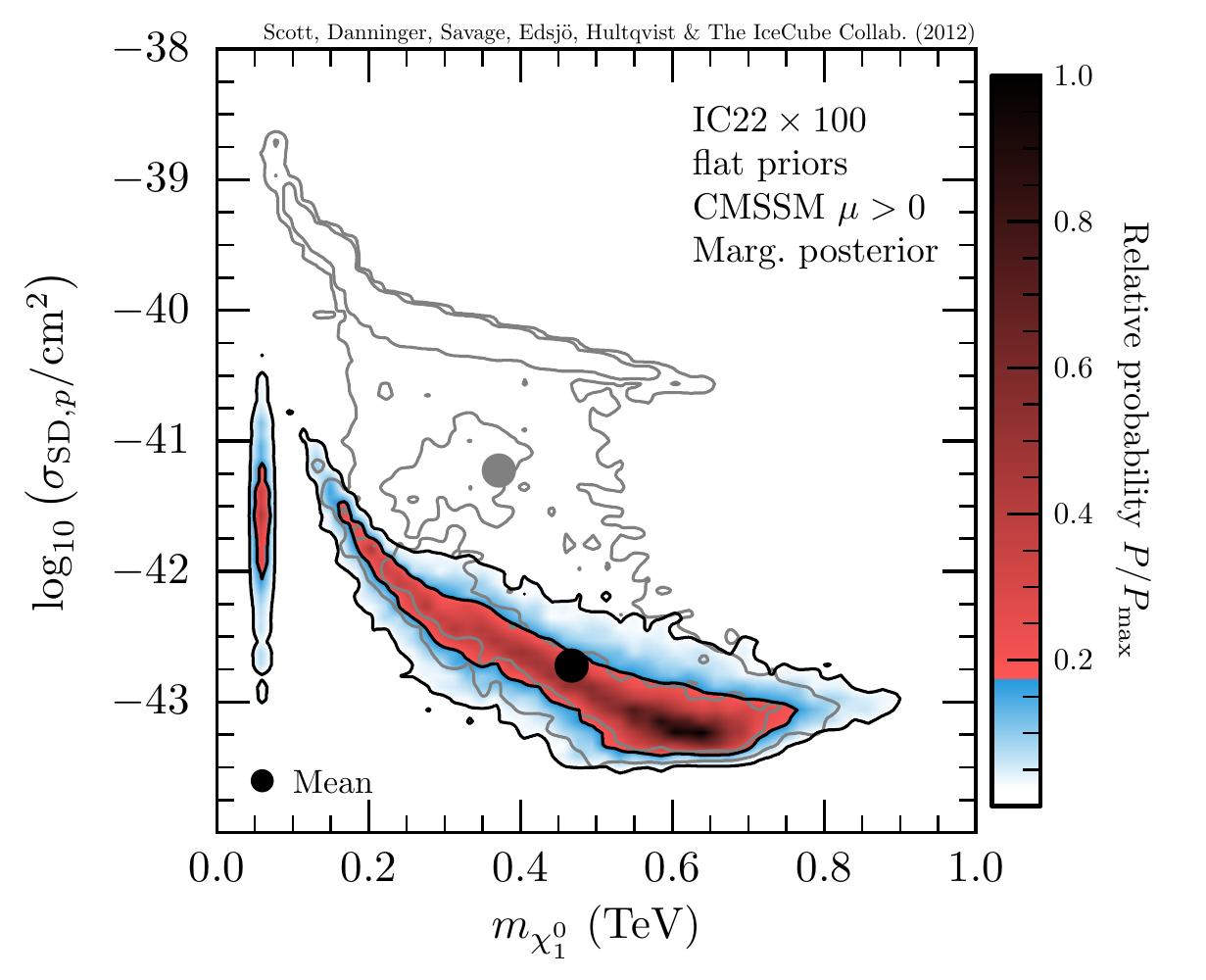}
\end{minipage}
\caption{As per Fig.~\protect\ref{fig:cmssmIC22}, but assuming that the detector effective area is boosted by a factor of 100, in order to make a rough estimate of the impact of the complete IceCube detector on the CMSSM.  Grey contours again correspond to the fit with no IceCube data.  Upcoming searches for dark matter with IceCube will robustly exclude the majority of the focus point region of the CMSSM.}
\label{fig:cmssmIC22x100}
\end{figure}

As expected from the limits produced in the original 22-string analysis \cite{IceCube09}, this data has very little impact upon the preferred regions in a CMSSM fit; the stau co-annihilation region (the small region at low $m_0$ in the leftmost panel and the curved, lower cross-section regions in the other two panels) is unaffected, and any impact on the focus point region (the larger region at high $m_0$ in the leftmost panel and the nearly horizontal, high cross-section bands in the other panels) is difficult to see.  On very close comparison of the black and grey contours in the right panel of Fig.~\ref{fig:cmssmIC22} (somewhat easier if one actually refers to the grey contours in the right panel of the \textit{next} figure, Fig.~\ref{fig:cmssmIC22x100}, which are identical to the grey contours plotted in Fig.~\ref{fig:cmssmIC22}), a very small impact can perhaps be seen.  A tiny part of the focus point region, at large spin-dependent scattering cross-section and $m_{\chi_1^0}\sim200$\,GeV, appears to have been disfavoured by the IC22 likelihood at the level of moving it from the $1\sigma$ credible region to the $2\sigma$ region.  Whether this is simply scanning noise is difficult to say, but the fact that the region in question has a very high posterior PDF in the scan without IceCube data, and is therefore very well-sampled, would seem to argue against such an interpretation.

To illustrate the approximate effect on the CMSSM of an eventual non-detection of neutrinos from WIMP annihilation in the Sun by the full IceCube detector, using our likelihood formalism, we performed a second global fit with a rescaled IceCube effective area.  We multiplied the IC22 effective area by a factor of 100, and kept all other aspects of the detector as in the 22-string analysis (angular errors, event sample, backgrounds and spectral response); we refer to this as the `IC$22\times100$' configuration.  Although we employ the actual simulated 86-string analysis \cite{ICRC2011ic86} later for model exclusion, using it for a study such as we describe in this section is not possible, as the 86-string analysis does not contain the requisite $N_\mathrm{chan}$ information to include spectral information in the likelihood function.

The results of the IC$22\times100$ global fit are shown in Fig.~\ref{fig:cmssmIC22x100}.  Grey contours again refer to an identical scan performed without the inclusion of any IceCube data.  As is clear from these results, the sensitivity of something resembling the full IceCube detector to both spin-independent and spin-dependent WIMP-nucleon interactions should place very strong constraints on the CMSSM, all but ruling out the majority of the focus-point region.  Whilst this has been shown to already be the case when XENON-100 data is included in the global fit \cite{MastercodeXENON100,SuperBayeSXENON100}, those constraints are based only on spin-independent scattering, and are therefore particularly sensitive to the adopted prior on the hadronic matrix elements \cite{Ellis08}.  If the value of the pion nuclear sigma term is taken from lattice calculations instead of from experiment, XENON-100 provides almost no constraint on the focus point region.  IceCube should provide a more complete exclusion of most of the focus-point region than XENON-100, because both spin-independent and spin-dependent couplings are expected to contribute to the solar capture rate, and spin-dependent scattering is far less sensitive to hadronic uncertainties than spin-independent scattering.  The full IceCube result will therefore constitute an important independent verification of the XENON-100 exclusion; if IceCube sees a signal in its 86-string configuration, this will be a strong indication that the CMSSM is not responsible for dark matter, or that there is an error in the experimental determination of the pion nuclear sigma term.  Even if there is no signal, IceCube will have a major impact on the parameter space of less constrained versions of supersymmetry, where the spin-independent and spin-dependent nuclear scattering cross-sections are not so tightly coupled as in the CMSSM.

The appearance of the narrow region at low $m_{1/2}$ and $m_{\chi_1^0}$ (the well-known CMSSM light Higgs funnel region) only in the IC$22\times100$ scan is not surprising, and can be understood via a combination of prior and scanning effects.  This region `exists' as a locale of reasonable fits in all scans, but only appears in the IC$22\times100$ posterior plots because in this case, given the linear prior we have employed on $m_{1/2}$ and $m_0$, removing most of the focus point increases the relative contribution of the funnel to the final model evidence.  These nuances are not directly related to the inclusion of IceCube data, and have been examined extensively already in the literature \cite{Trotta08,Akrami09,SBSpike}.

\begin{figure}[p]
\vspace{-5mm}
\centering
\begin{minipage}[b]{0.48\textwidth}
a)
\end{minipage}
\hspace{0.01\textwidth}
\begin{minipage}[b]{0.48\textwidth}
b)
\end{minipage}
\begin{minipage}[b]{0.48\textwidth}
\centering
\includegraphics[width=\linewidth]{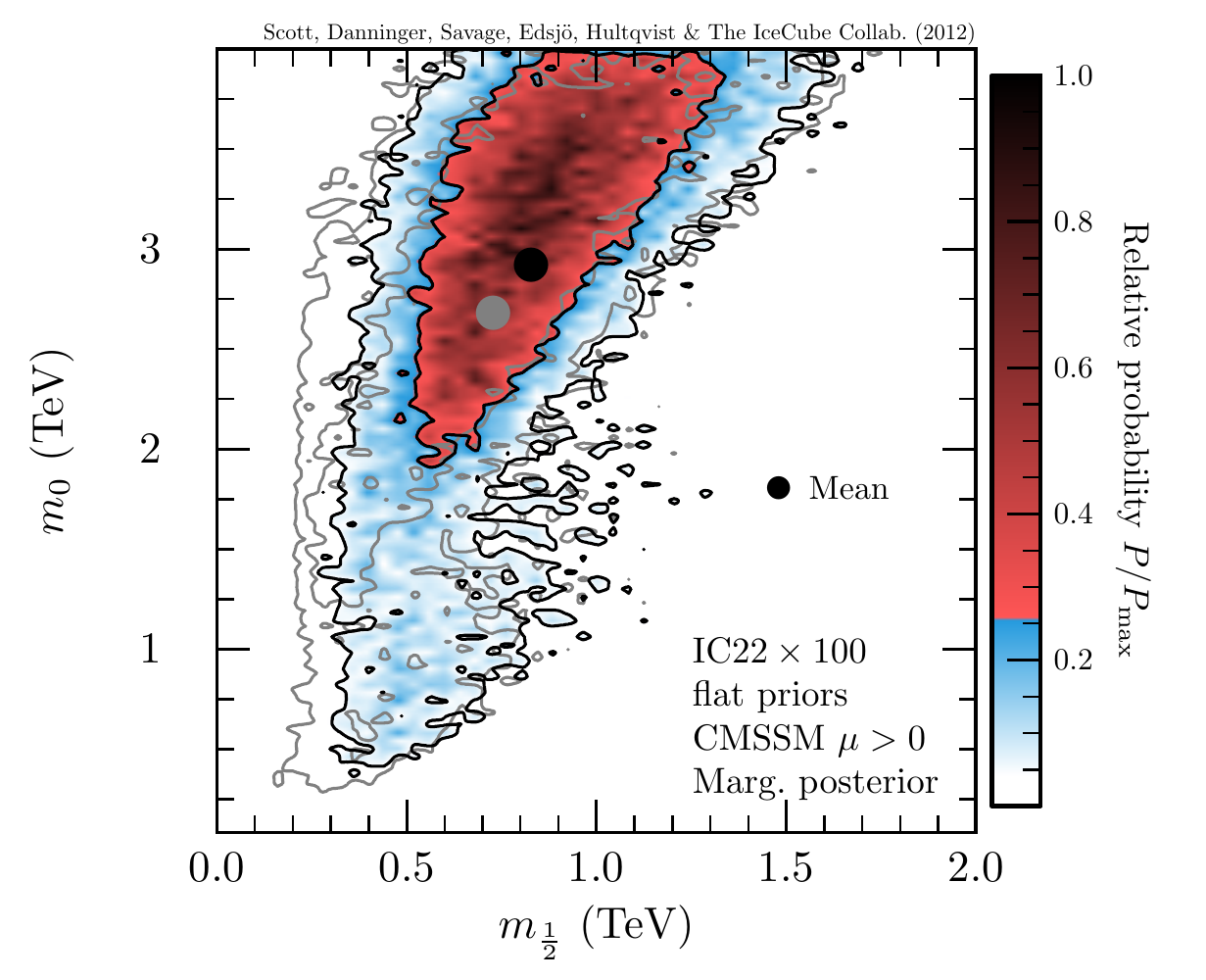}
\end{minipage}
\hspace{0.01\textwidth}
\begin{minipage}[b]{0.48\textwidth}
\centering
\includegraphics[width=\linewidth]{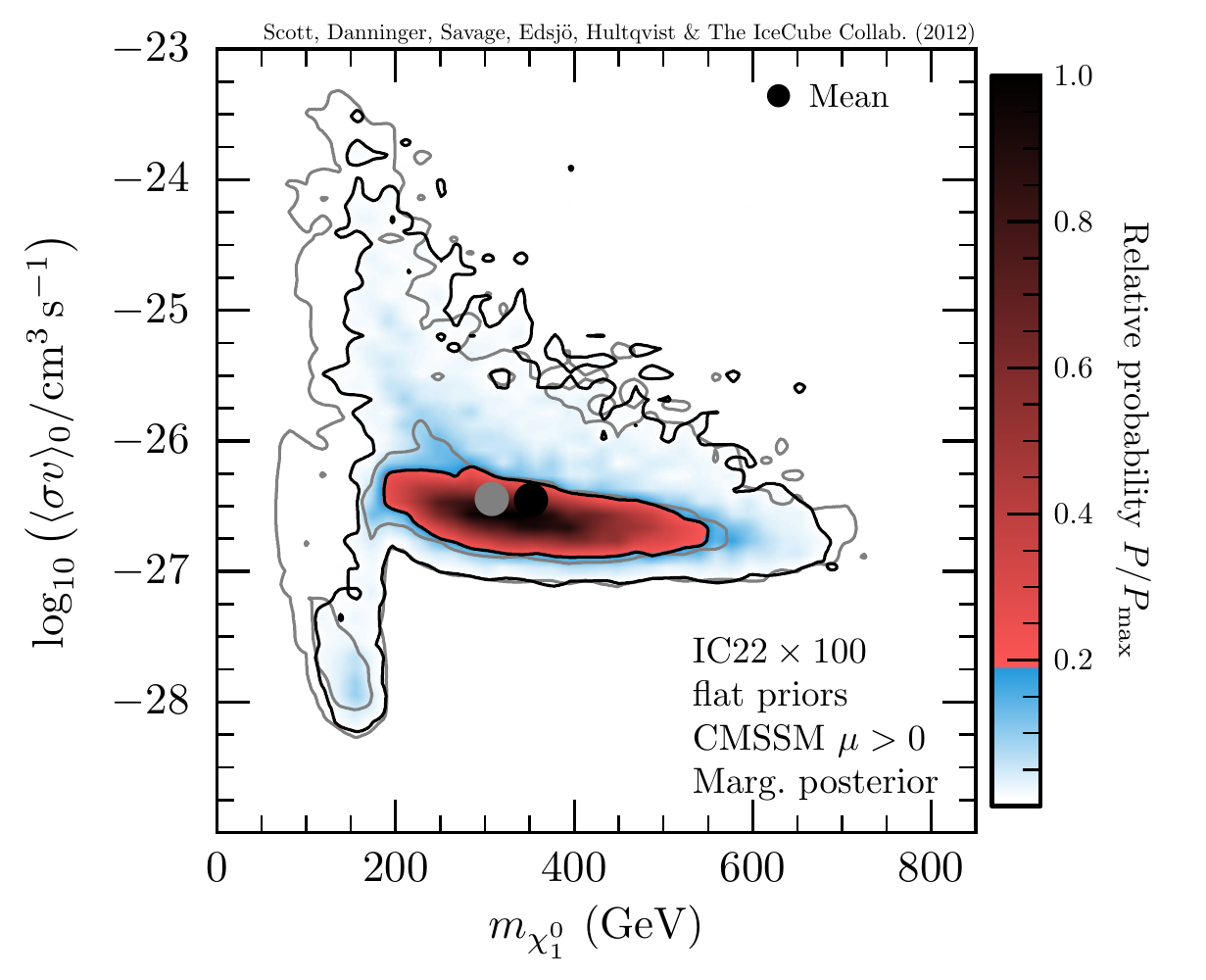}
\end{minipage}
\begin{minipage}[b]{0.48\textwidth}
c)
\end{minipage}
\hspace{0.01\textwidth}
\begin{minipage}[b]{0.48\textwidth}
d)
\end{minipage}
\begin{minipage}[b]{0.48\textwidth}
\centering
\includegraphics[width=\linewidth]{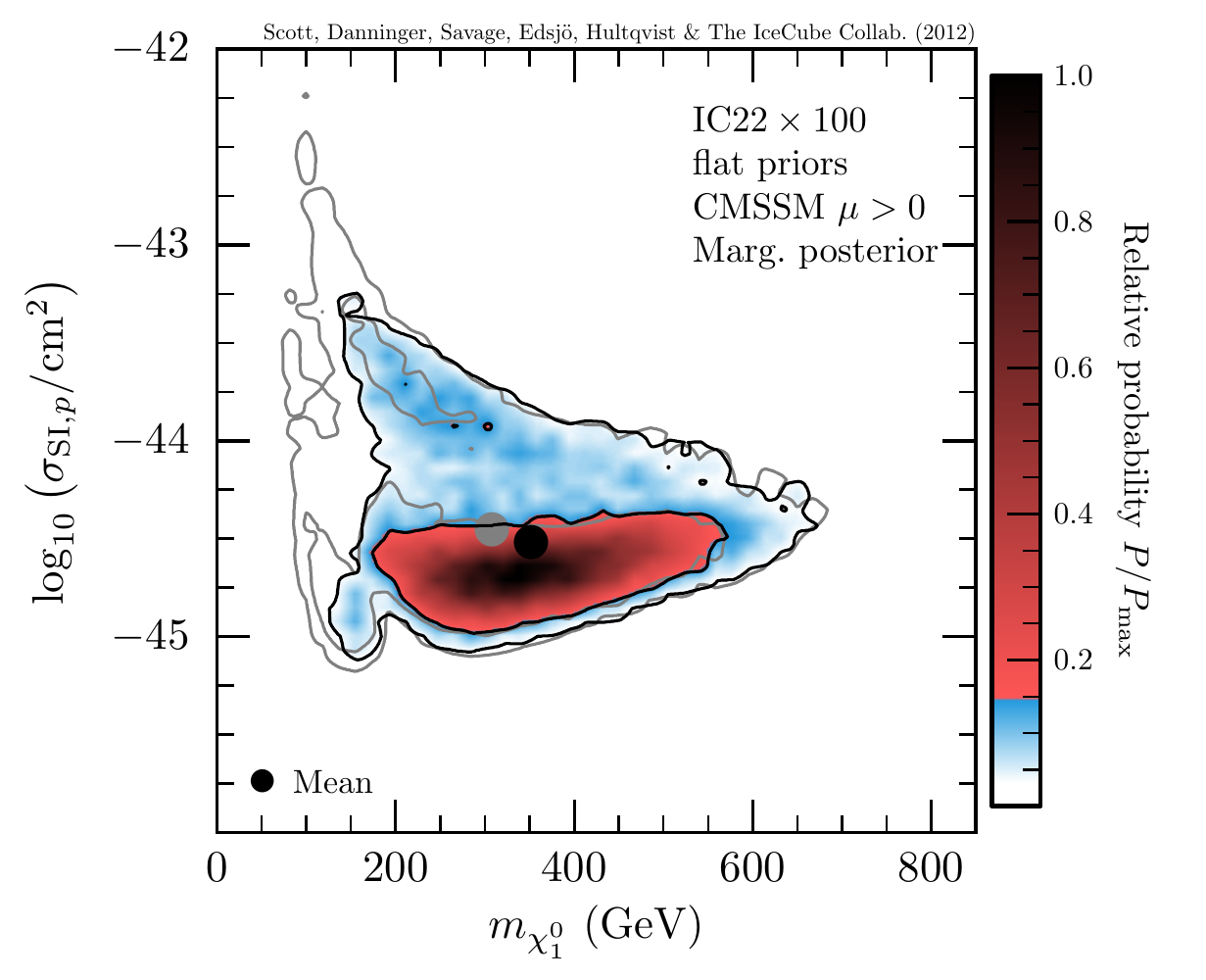}
\end{minipage}
\hspace{0.01\textwidth}
\begin{minipage}[b]{0.48\textwidth}
\centering
\includegraphics[width=\linewidth]{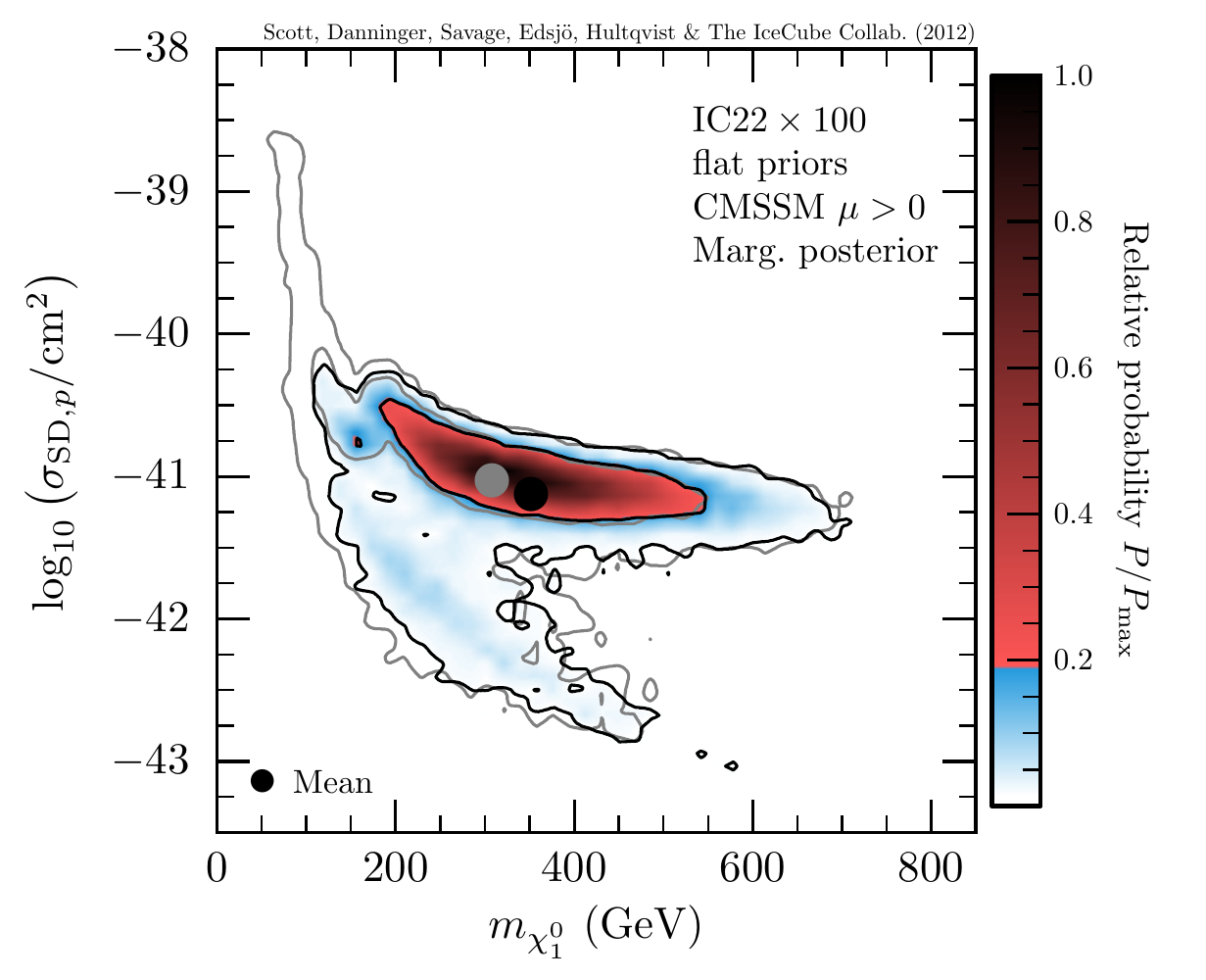}
\end{minipage}
\begin{minipage}[b]{0.48\textwidth}
e)
\end{minipage}
\hspace{0.01\textwidth}
\begin{minipage}[b]{0.48\textwidth}
f)
\end{minipage}
\begin{minipage}[b]{0.48\textwidth}
\centering
\includegraphics[width=\linewidth]{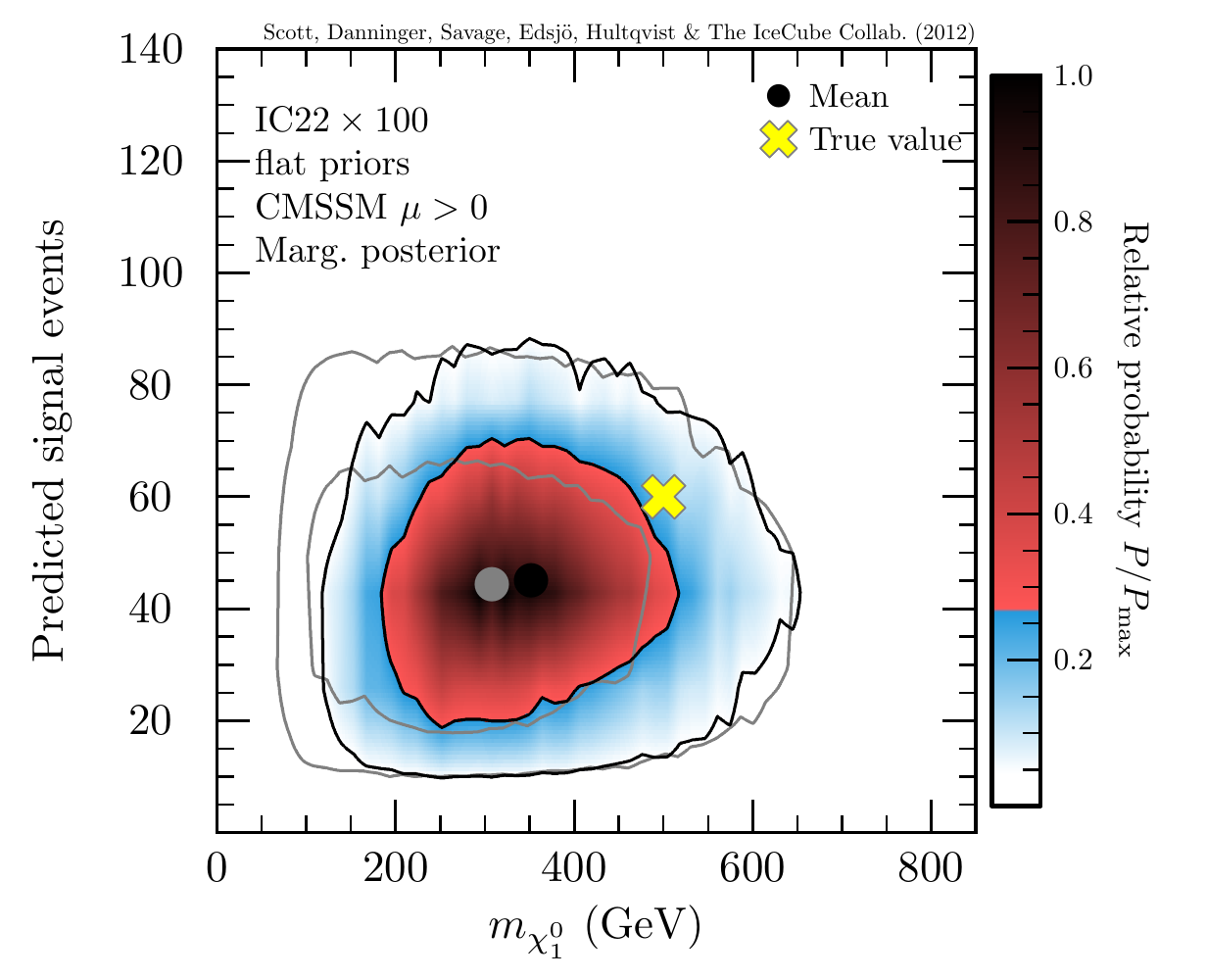}
\end{minipage}
\hspace{0.01\textwidth}
\begin{minipage}[b]{0.48\textwidth}
\centering
\includegraphics[width=\linewidth]{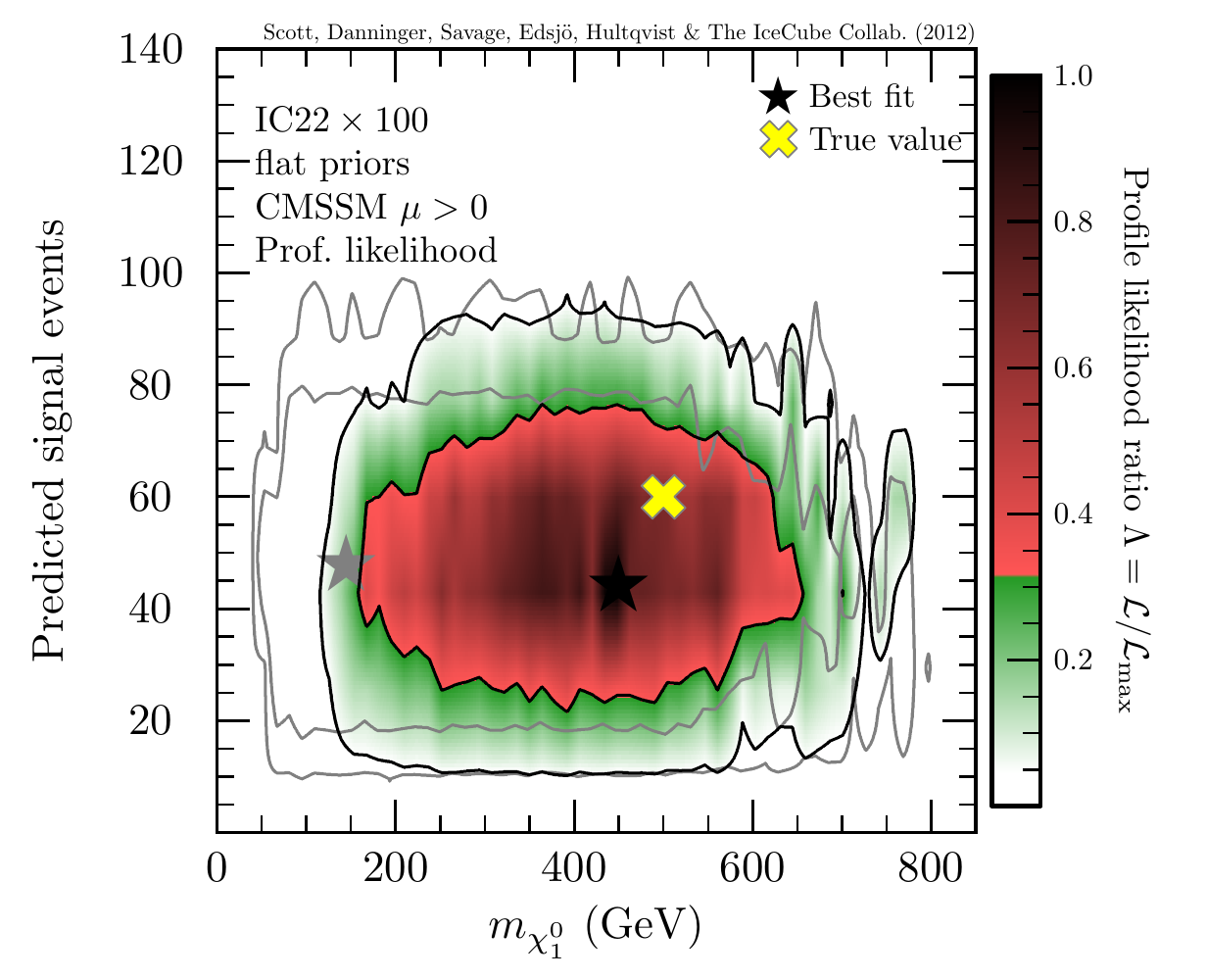}
\end{minipage}
\caption{2D posterior probability distributions (a -- e) and one comparison 2D profile likelihood (f), from a CMSSM signal recovery based on IceCube 22-string event data only.  We have again assumed that the detector effective area is boosted by a factor of 100 relative to the regular 22-string analysis, to approximate the final detector configuration of IceCube.  Here we have injected a simulated signal corresponding to a 500\,GeV WIMP annihilating into $W^+W^-$, at a rate that would give 60 signal events on average.  Contours indicate 68.3\% ($1\sigma$) and 95.4\% ($2\sigma$) credible/confidence regions.  Shading and black contours indicate the recovery using the full likelihood (Eq.~\ref{full_final_unbinned_like}), containing number, angular and spectral information.  Grey contours give the recovery achieved with only number and angular information.  The recovery is quite tight, especially with the inclusion of spectral information.}
\label{fig:cmssmrecon}
\end{figure}
\afterpage{\clearpage}

\subsection{CMSSM recovery validation with mock IceCube data}
\label{cmssmrecon}

To explicitly test the validity and performance of the likelihood construction outlined in Sec.~\ref{likelihood}, we performed a signal recovery within the CMSSM, using the IC$22\times100$ detector configuration.  We simulated a WIMP annihilation signal in the Sun expected to produce, on average, 60 signal events within the 10 degree angular cut.  We did this as a single realisation of the angular and spectral distribution expected from a 500\,GeV WIMP annihilating exclusively into $W^+W^-$ pairs, taking into account the angular resolution of the 22-string detector and the expected distribution of $N_\mathrm{chan}$ values from such a signal.  We added this signal to a realisation of the expected background, which we obtained by bootstrap Monte Carlo simulation using the actual all-sky events observed in the 22-string analysis (as described in Sec.~\ref{icbackground}).  The particular realisation we employed contained 53 signal and 164 background events within 10 degrees of the solar position.

We then ran a full parameter scan based on the simulated event list, using \iclike\ and our modified version of \textsf{SuperBayeS}.  For this recovery we included \textit{only} the simulated IceCube data in the likelihood function (i.e.\ no relic density or other experimental constraints), as we are specifically interested in the ability of IceCube to pin down parameters of a theory of new physics using our likelihood formalism.

\begin{figure}[t]
\centering
\begin{minipage}[b]{0.48\textwidth}
a)
\end{minipage}
\hspace{0.01\textwidth}
\begin{minipage}[b]{0.48\textwidth}
b)
\end{minipage}
\begin{minipage}[b]{0.48\textwidth}
\centering
\includegraphics[width=\linewidth]{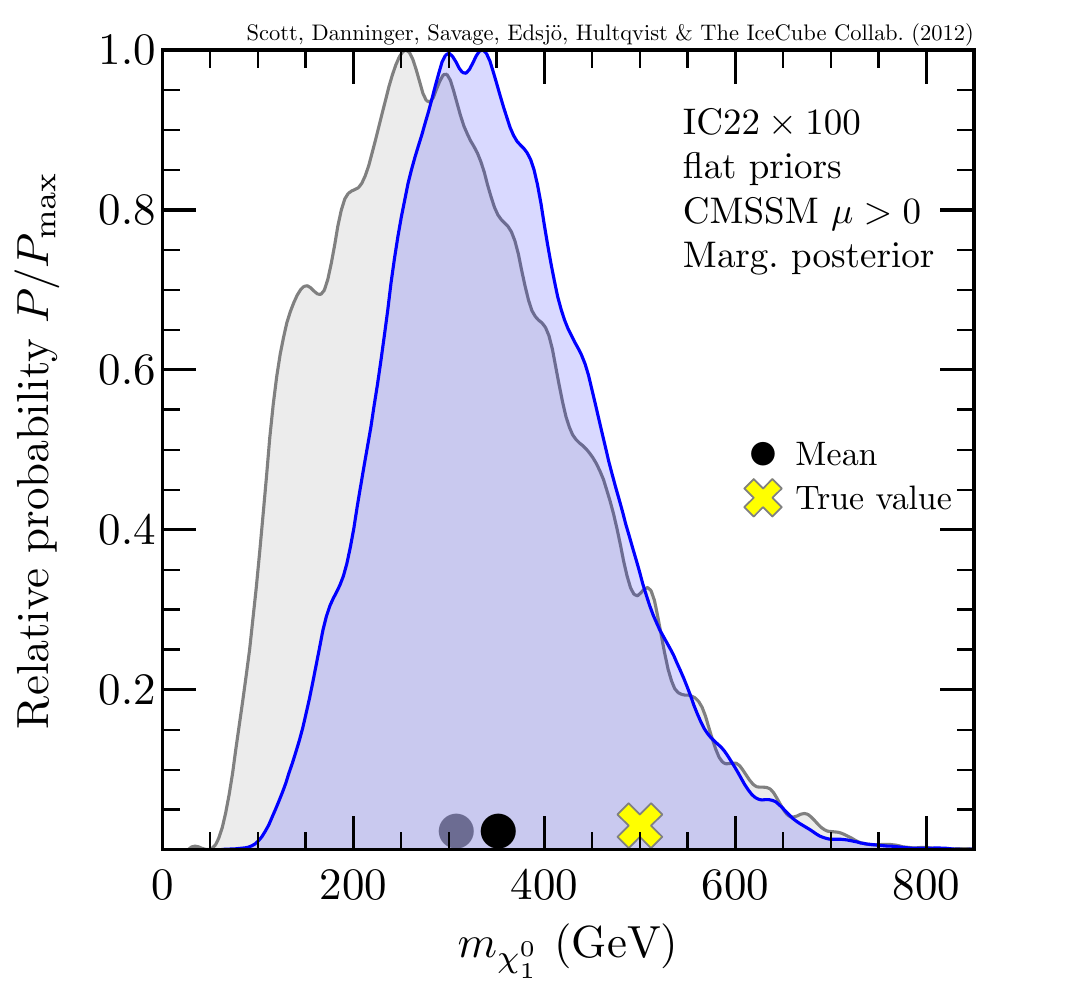}
\end{minipage}
\hspace{0.01\textwidth}
\begin{minipage}[b]{0.48\textwidth}
\centering
\includegraphics[width=\linewidth]{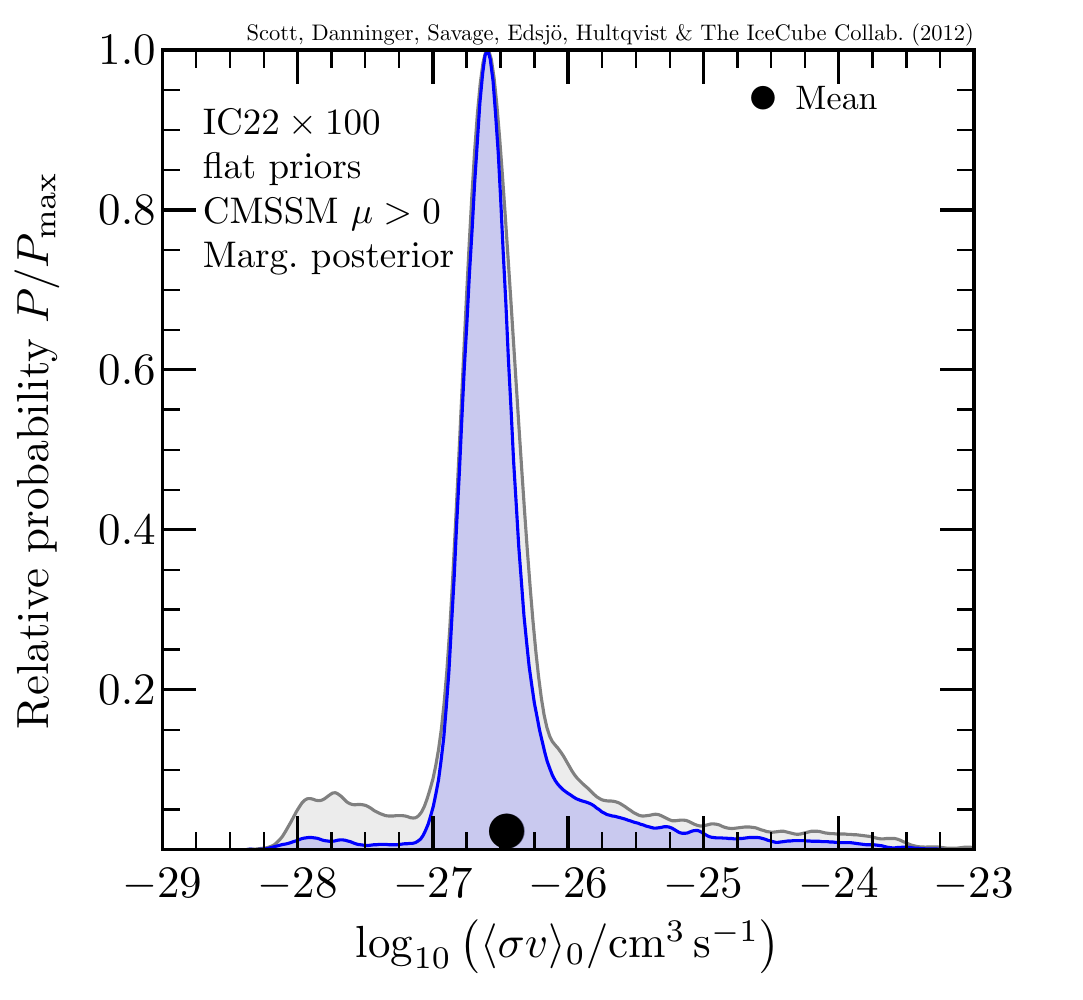}
\end{minipage}
\begin{minipage}[b]{0.48\textwidth}
c)
\end{minipage}
\hspace{0.01\textwidth}
\begin{minipage}[b]{0.48\textwidth}
d)
\end{minipage}
\begin{minipage}[b]{0.48\textwidth}
\centering
\includegraphics[width=\linewidth]{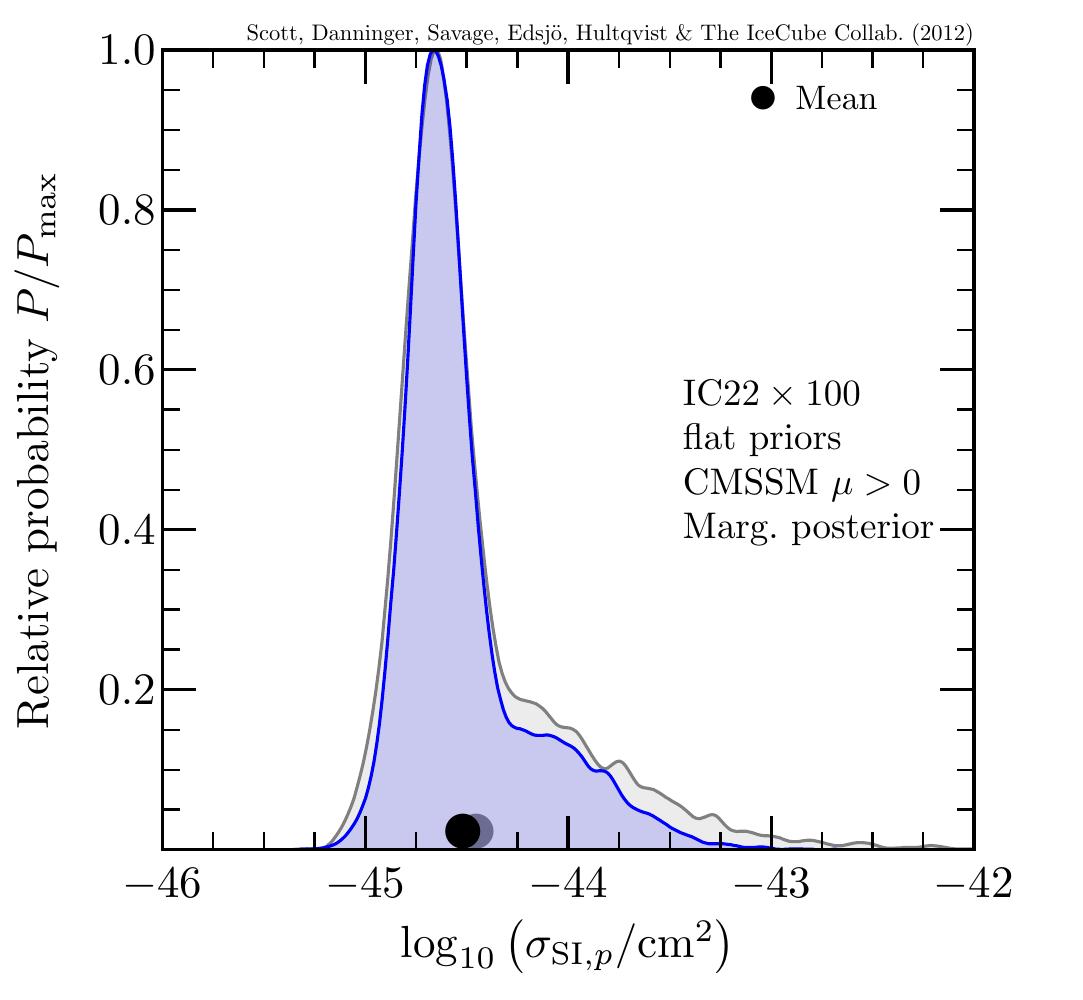}
\end{minipage}
\hspace{0.01\textwidth}
\begin{minipage}[b]{0.48\textwidth}
\centering
\includegraphics[width=\linewidth]{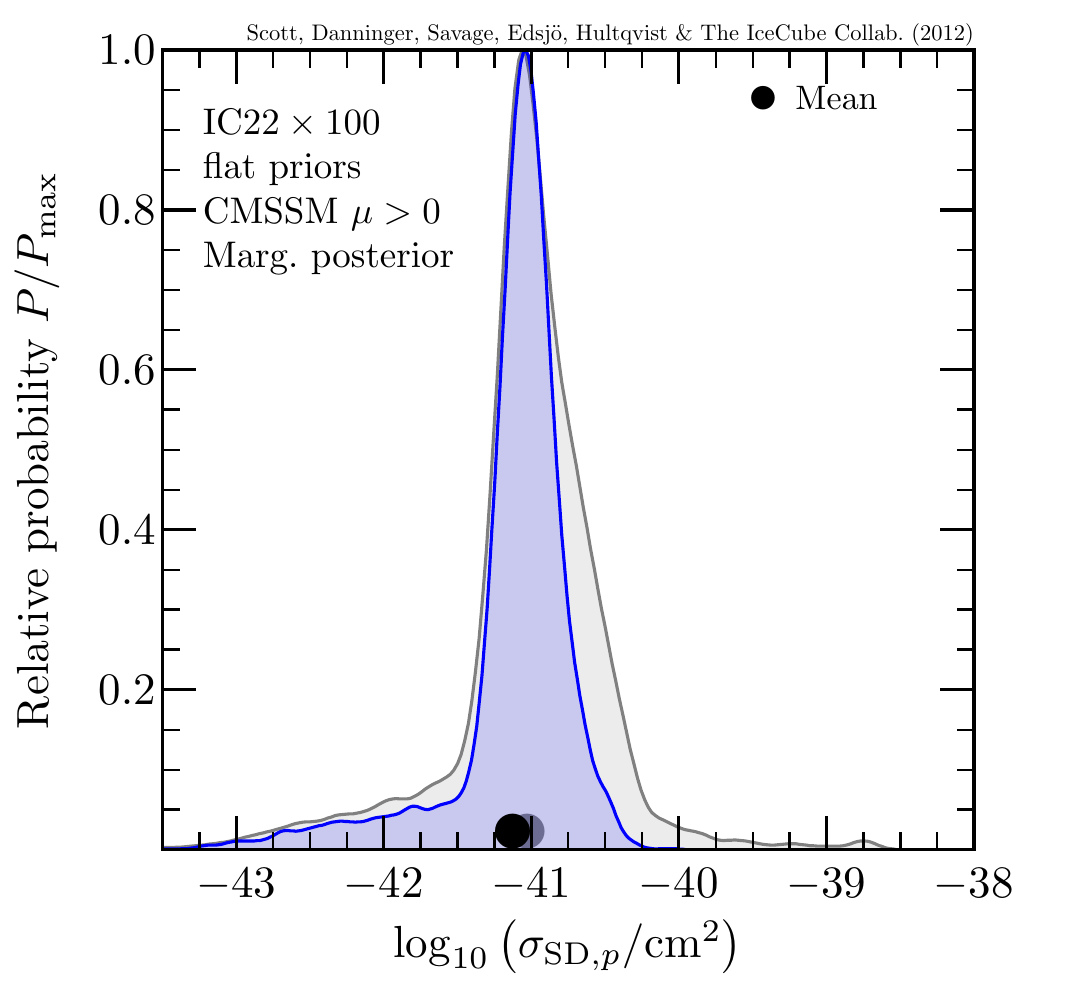}
\end{minipage}
\caption{1D posterior probability distributions for the mass of the lightest neutralino and its most important cross-sections, from a CMSSM signal recovery.  This is the same recovery shown in Fig.~\protect\ref{fig:cmssmrecon}, based on IceCube 22-string event data only, but assuming a factor of 100 boost in the effective area in order to approximate the final detector configuration.  Blue curves give the probability distributions obtained using the full likelihood (Eq.~\ref{full_final_unbinned_like}).  Grey curves show the distributions achieved without the use of spectral information.  The use of spectral information significantly improves the accuracy with which the mass and cross-sections can be determined.}
\label{fig:cmssmrecon1D}
\end{figure}

In Fig.~\ref{fig:cmssmrecon} we show the results of the recovery in terms of various two-dimensional marginalised posterior PDFs.  Here we give not only joint distributions for $m_0$ and $m_{1/2}$, and for $m_{\chi_1^0}$ and the two nuclear-scattering cross-sections, but also for $m_{\chi_1^0}$ and the velocity-averaged annihilation cross-section $\langle \sigma v\rangle_0$, and $m_{\chi_1^0}$ and the number of predicted signal events in IC$22\times100$.  We also give in Fig.~\ref{fig:cmssmrecon1D} the 1D marginalised posterior probability distributions of the WIMP mass $m_{\chi_1^0}$, and the three cross-sections.  For comparison, we give the results of a similar recovery \textit{not including the spectral component of the likelihood} in grey.

Figs.~\ref{fig:cmssmrecon} and \ref{fig:cmssmrecon1D} show that that our technique accurately recovers both the WIMP mass and expected number of events, and results in rather accurate measurements of the various cross-sections.  Fig.~\ref{fig:cmssmrecon}a demonstrates that in this particular example, the fit hones in quite effectively on a part of the focus point region, corresponding to quite small regions in the mass--cross-section planes (Figs.~\ref{fig:cmssmrecon}b,c,d).  The performance in the cross-section directions (Figs.~\ref{fig:cmssmrecon}b,c,d and \ref{fig:cmssmrecon1D}b,c,d) is particularly impressive given that our benchmark point does not specify any cross-sections, only an expected number of events.  A number of different combinations of cross-sections result in an appropriate number of events, as capture and annihilation need not be in equilibrium in the Sun in our calculation, and the relative contributions of spin-independent and spin-dependent scattering to the total capture rate is not specified.  

Although the benchmark point lies just outside the edge of the $1\sigma$ posterior credible region in $m_{\chi_1^0}$ and the predicted number of signal events (Fig.~\ref{fig:cmssmrecon}e), this is entirely consistent with realisation noise and the impact of priors. The profiled likelihood function (Fig.~\ref{fig:cmssmrecon}f) in this plane easily encompasses the true value within its $1\sigma$ confidence interval, indicating that the slight preference towards lower masses in the posterior PDF is a prior-driven effect.  As expected, Fig.~\ref{fig:cmssmrecon}f is somewhat noisy because the scan was not optimised for profile likelihoods, but the convergence towards a roughly Gaussian likelihood in this plane is obvious.  Recall also that the benchmark point refers to the expected number of total signal events (60), but that the random realisation that we happen to employ contains 53 signal events inside the analysis cone.  It is therefore not at all surprising to find that the best-fit and posterior mean values returned by the scan lie in the range 45--50, given the size of the credible/confidence regions.

The grey contours in Fig.~\ref{fig:cmssmrecon}, and corresponding grey curves in Fig.~\ref{fig:cmssmrecon1D}, illustrate the dramatic improvement in the mass recovery achieved by including even a crude energy estimator like $N_\mathrm{chan}$.  We see that the full likelihood consistently disfavours low masses compared to the recovery without spectral information.  This results in smaller credible regions in all planes of Fig.~\ref{fig:cmssmrecon}, particularly those involving cross-sections, and tighter credible intervals in Fig.~\ref{fig:cmssmrecon1D}.  Unsurprisingly, excluding the angular part of the likelihood also has a dramatic effect; using just the number likelihood results in a $2\sigma$ credible region that covers most of the parameter space.

\begin{figure}[t]
\centering
\hspace{-0.07\textwidth}
\begin{minipage}[t]{0.31\textwidth}
\centering
\includegraphics[height=1.1\linewidth]{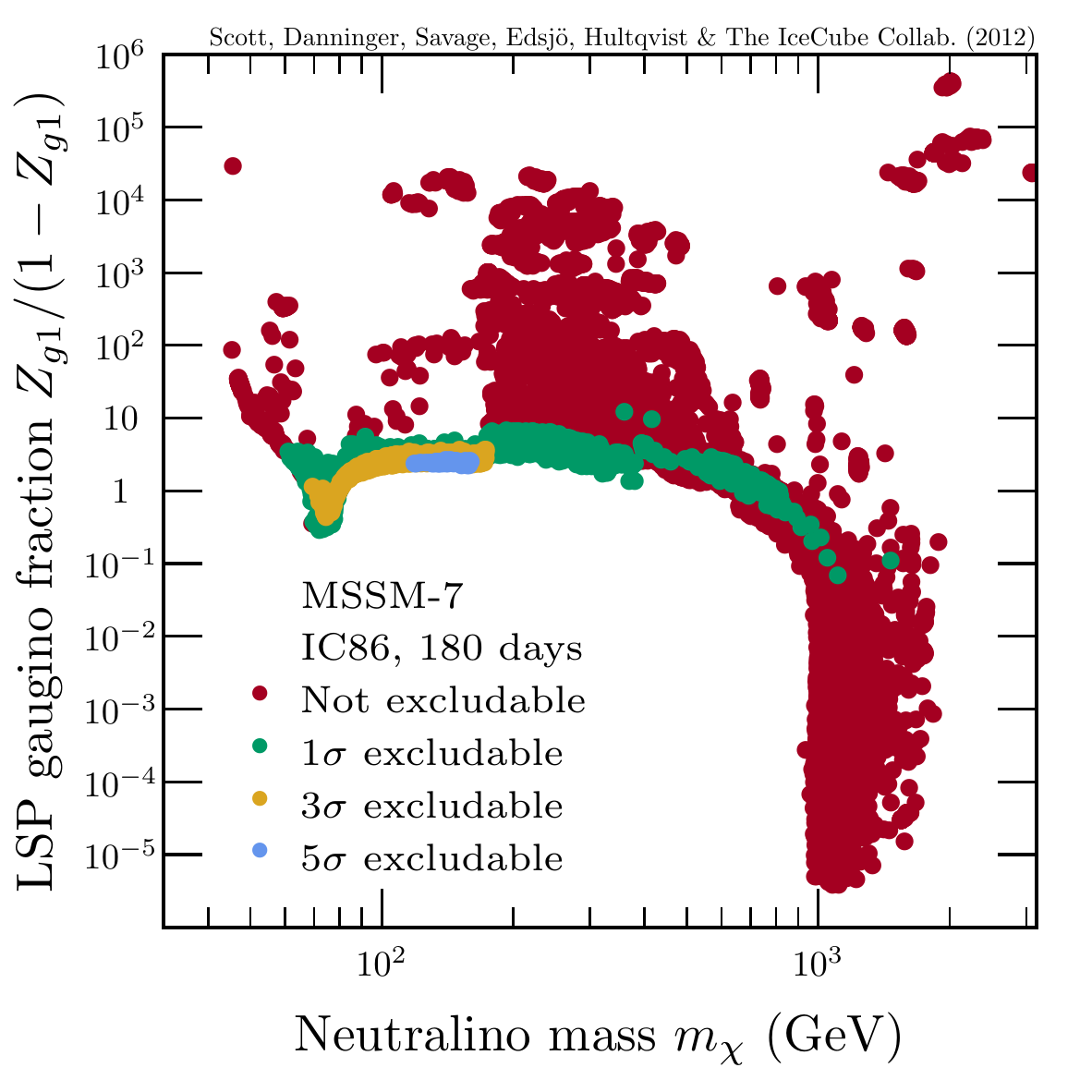}
\end{minipage}
\hspace{0.03\textwidth}
\begin{minipage}[t]{0.31\textwidth}
\centering
\includegraphics[height=1.1\linewidth]{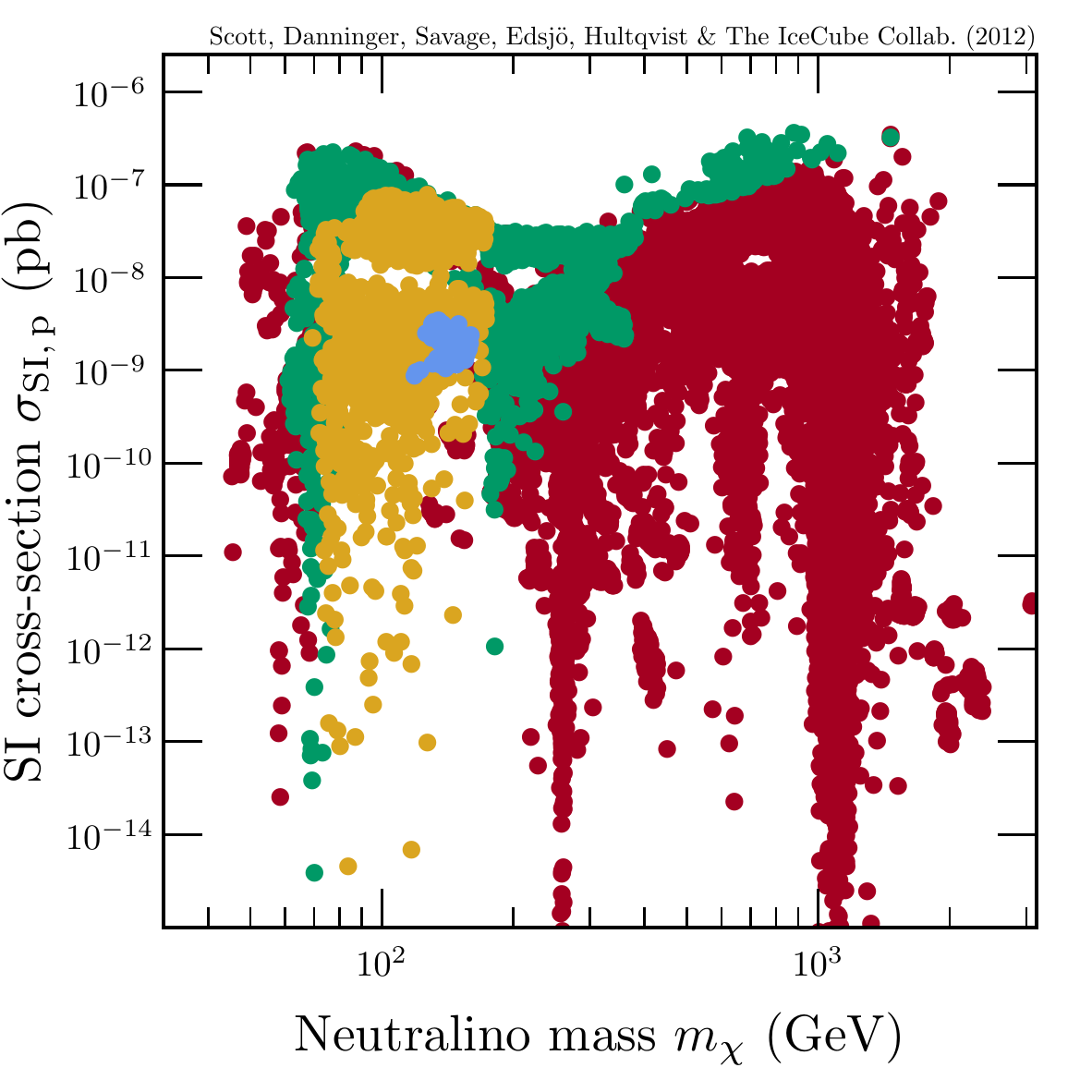}
\end{minipage}
\hspace{0.03\textwidth}
\begin{minipage}[t]{0.31\textwidth}
\centering
\includegraphics[height=1.1\linewidth]{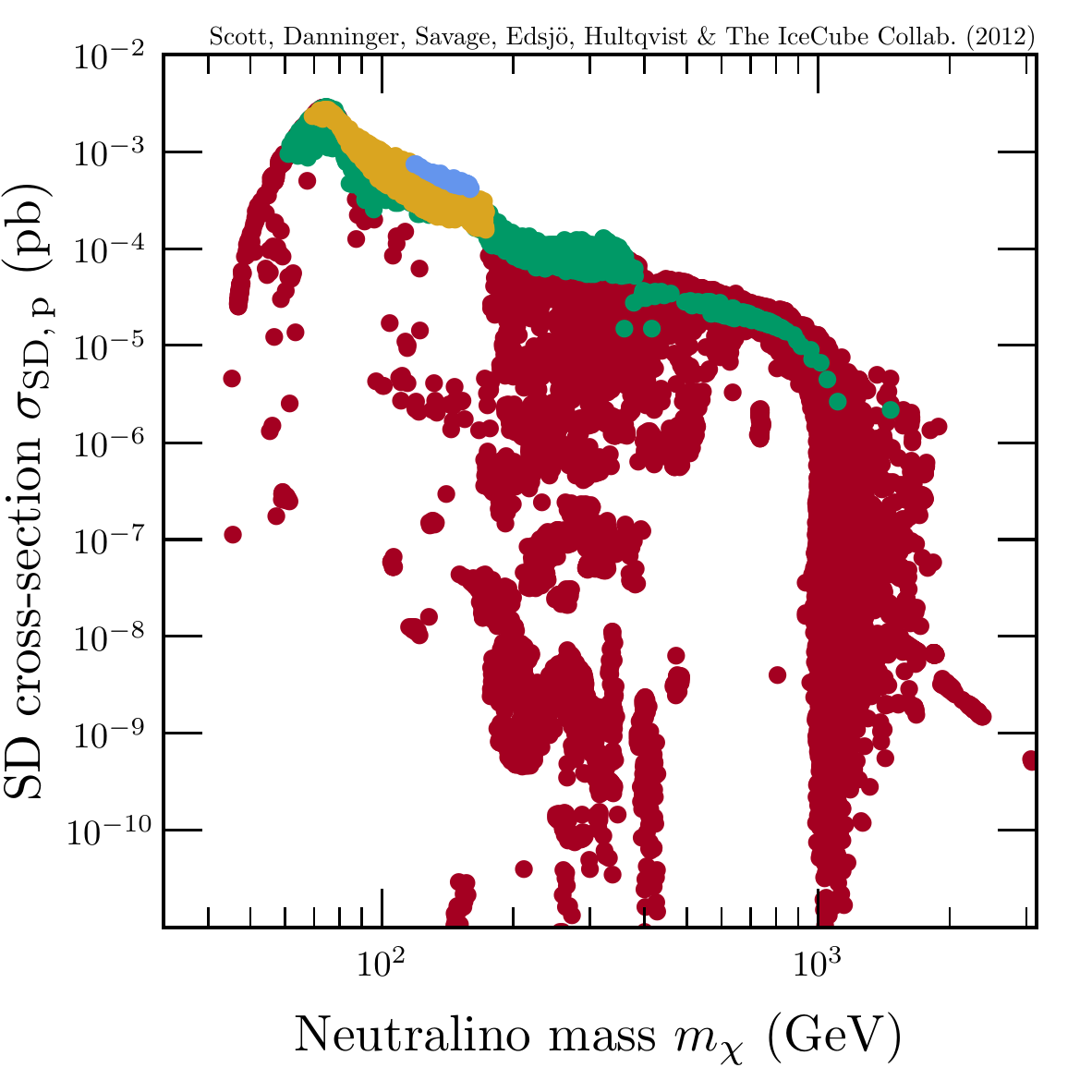}
\end{minipage}
\caption{A model exclusion analysis in the MSSM-7, based on Eq.~\protect\ref{finalpval} and the consideration of all events seen to arrive from a direction within 20 degrees of the solar position.  Here we use a simulated 86-string IceCube configuration, and simulated data consistent with background.  A number of models will be directly excludable by the full IceCube detector, without any recourse to the rest of the model parameter space, nor any spectral or angular information beyond the number of events within the angular cut cone.  The strongest exclusions will be for models with mixed gaugino/Higgsino dark matter, giving rise to large spin-dependent nuclear scattering rates.  In particular, many of these models have spin-independent cross-sections too low to be detected by large terrestrial experiments like XENON-100.}
\label{fig:mssm7}
\end{figure}

\subsection{MSSM-7 model exclusion with 86-string simulation}
\label{mssm7}

We have also performed scans in the more phenomenological MSSM-7 with seven free parameters at the electroweak scale: the Higgsino mass parameter $\mu$, the gaugino mass parameter $M_2$, the ratio of the Higgs vacuum expectation values $\tan \beta$, a common sfermion mass parameter $m_0$, the CP-odd Higgs boson mass $m_A$ and the trilinear couplings for the third generation squarks $A_b$ and $A_t$. See e.g.\ \cite{BergstromGondolo96} for a more detailed description of this model. 

The database we use here is made from several scans focusing on different regions of the parameter space. The core of the database consists of general random scans allowing the mass parameters to take on values up to 10\,TeV, $A_t$ and $A_b$ in the range between $-3m_0$ and $3m_0$ and $\tan \beta$ between 1.2 and 60. We have also performed importance sampling scans with ADSCAN \cite{Brein:2004kh}. ADSCAN is setup to use an adaptive integration routine (VEGAS) based on importance sampling. We have then chosen to run ADSCAN with different ``importance functions'' $G$ focusing on different phenomenologically interesting areas in the parameter space. The importance function we have mostly used is
$$
G\left(\Omega_\chi h^2\right) = \exp\left[-\frac12\left(\frac{\Omega_\chi h^2-\Omega_{\rm WMAP}h^2}{\sigma_{\Omega h^2}}\right)^2\right],
$$
with $\Omega_{\rm WMAP}h^2 = 0.1099$ from WMAP data \cite{wmap5}. For the error we have chosen $\sigma_{\Omega h^2}=0.01$ to allow for some theoretical error on top of the experimental one. We have also chosen importance functions where we add factors to drive the scans into regions of high gaugino fractions, high scattering cross sections or low neutralino masses. To get a well-sampled database we have also forced ADSCAN into certain mass ranges with importance functions of the type
$$
G\left(\Omega_\chi h^2, m_{\chi_1^0}\right) = \exp\left[-\frac12\left(\frac{\Omega_\chi h^2-\Omega_{\rm WMAP}h^2}{\sigma_{\Omega h^2}}\right)^2 - \frac12\left(\frac{m_{\chi_1^0}-m_{\rm target}}{\Delta m}\right)^2 \right],
$$
varying $m_{\rm target}$ between 10\,GeV and several\,TeV, and commonly using $\Delta m=20$\,GeV.

In Fig.~\ref{fig:mssm7} we show the results of a model exclusion analysis performed on this database for an 86-string IceCube configuration (IC86).

The detector will be most sensitive to the region where the neutralino is an almost equal Higgsino-gaugino mixture, which is where the spin-dependent scattering cross-section is usually the largest (as can be seen especially in the right panel in Fig.~\ref{fig:mssm7}).  This occurs because the spin-dependent cross-section is dominated by $t$-channel $Z$ exchange, which couples to the difference between the two Higgsino contributions; if the neutralino is strongly gaugino, this coupling is absent, whereas if it is strongly Higgsino it becomes equal parts $\tilde{H}_1^0$ and $\tilde{H}_2^0$ and the two contributions cancel.

As many of these models are captured in the Sun mainly via spin-dependent scattering, the current direct detection experiments (which are mostly sensitive to the spin-independent scattering cross section) do not probe most of the models to which IC86 will be most sensitive. This can be seen in the middle panel, where many of the models IceCube can probe are far below the sensitivities of both current and near-future direct detection experiments.

Compared to the standard IceCube `hard' and `soft'-channel analyses \cite{ICRC2011ic86}, which are valid for just two specific annihilation final states, the (mixed final state) MSSM-7 models excluded in the analysis here lie (predictably) between the hard and soft sensitivity curves.  The implied sensitivity is closer to the soft-channel curve than it would have been if we had employed a similarly aggressive angular cut as in \cite{ICRC2011ic86}.  For this analysis we used an angular cut of $\phi_\mathrm{cut}=20^\circ$ around the solar position.  In general we have chosen angular cuts so as to include the majority of a potential signal from the Sun.  The cut angle may be optimised for different models however, as higher-mass WIMPs would certainly produce more centrally-localised, high-energy neutrino events.  This is particularly important when doing model exclusion, as the $p$ value Eq.~\ref{finalpval} depends only on the expected number of signal and background events, not their angular distributions.  The exclusion power of the full likelihood Eq.~\ref{full_final_unbinned_like} is less sensitive to the cut angle, but is still not independent of it, because of the necessary approximations we had to make about the separability of the spectral and angular components.  In principle the cut angle could be profiled or marginalised over as a nuisance parameter in model scans, but this is probably overly conservative; it can be optimised and chosen in advance for specific models based on simulated data, and an approximate scheme (depending, say, on WIMP mass and annihilation branching fractions) constructed for choosing the approximate optimal cut angle for any arbitrary model.  Such an extension may be implemented in future updates to the formalism we have presented in this paper.

\section{Conclusions}
\label{conclusions}

We have constructed an unbinned likelihood formalism for including full event-level IceCube data in parameter explorations of theories for new physics.  The likelihood function includes information about the number, direction and spectral characteristics of neutrino events, and is fast to calculate. We also constructed a simple associated measure for model exclusion, which is even faster to calculate and can be used for single models, without any reference to other parts of a parameter space.

We performed a number of example model scans using our likelihood construction within the MSSM.  We carried out global fits to the CMSSM with actual 22-string data, and with the 22-string effective area rescaled to represent the final detector configuration.  Existing 22-string data has little impact on the CMSSM, but the final detector configuration will robustly exclude the majority of the focus point region if IceCube finds no evidence for WIMP annihilation.  We carried out a mock signal recovery in the CMSSM, showing that our method accurately recovers a benchmark point, and constrains model parameters very well.  In the process, we showed the utility of including spectral information in IceCube searches for dark matter.  Finally, we carried out a simple example model-exclusion analysis in the MSSM-7, showing that the 86-string configuration of IceCube will test a number of models that cannot be probed by direct detection experiments in the near future.

Our likelihood construction is implemented in \iclike, and freely accessible to the community.  It will also be made available in a future release of \textsf{SuperBayeS}.  The data and simulations we have used for this study, including 22-string IceCube event lists \cite{IceCube09}, are freely available on the web \cite{filesloc}, and in \iclike.

\acknowledgments{We are grateful to Hamish Silverwood for helpful comments and discussions.  P.S. is supported by the Lorne Trottier Chair in Astrophysics and a Canadian Institute for Particle Physics Theory Fellowship.  M.D. and K.H. acknowledge support from the Swedish Research Council (Contract No.\ 621-2010-3705), as does J.E. (Contract No.\  621-2010-3301).

We acknowledge the support from the following agencies:
U.S. National Science Foundation -- Office of Polar Programs,
U.S. National Science Foundation -- Physics Division,
University of Wisconsin Alumni Research Foundation,
the Grid Laboratory Of Wisconsin (GLOW) grid infrastructure at the University of Wisconsin - Madison, the Open Science Grid (OSG) grid infrastructure;
U.S. Department of Energy, and National Energy Research Scientific Computing Center,
the Louisiana Optical Network Initiative (LONI) grid computing resources;
National Science and Engineering Research Council of Canada;
Swedish Research Council,
Swedish Polar Research Secretariat,
Swedish National Infrastructure for Computing (SNIC),
and Knut and Alice Wallenberg Foundation, Sweden;
German Ministry for Education and Research (BMBF),
Deutsche Forschungsgemeinschaft (DFG),
Research Department of Plasmas with Complex Interactions (Bochum), Germany;
Fund for Scientific Research (FNRS-FWO),
FWO Odysseus programme,
Flanders Institute to encourage scientific and technological research in industry (IWT),
Belgian Federal Science Policy Office (Belspo);
University of Oxford, United Kingdom;
Marsden Fund, New Zealand;
Australian Research Council;
Japan Society for Promotion of Science (JSPS);
the Swiss National Science Foundation (SNSF), Switzerland.}

\bibliography{DMbiblio,SUSYbiblio}

\end{document}